\documentclass[twocolumn]{aastex631}

\usepackage{newtxtext,newtxmath}
\usepackage{amsmath}
\usepackage{latexsym}
\usepackage{amssymb}	
\usepackage{gensymb}
\usepackage{flafter}
\usepackage{appendix}
\usepackage{natbib}
\graphicspath{{./}{figures/}}

\usepackage{subfigure}
\usepackage[figuresright]{rotating}
\usepackage{makecell}
\usepackage{booktabs}
\usepackage{multirow}
\usepackage{threeparttable}
\usepackage{color}
\newcommand\hxmt{{\it Insight}-HXMT}

\received{Jan 18, 2022}
\revised{Feb 17, 2022 and May 29, 2022}
\accepted{June 12, 2022}
\submitjournal{ApJ}

\begin{document}
\title{Evolution of Accretion Modes between Spectral States Inferred from Spectral and Timing Analysis of Cygnus X-1 with \textit{Insight}{\rm -HXMT} observations}

\correspondingauthor{M. Z. , Feng}
\email{fengmz@ihep.ac.cn}
\correspondingauthor{L. D. , Kong}
\email{kongld@ihep.ac.cn}
\correspondingauthor{P. J. , Wang}
\email{wangpj@ihep.ac.cn}

\author{M. Z. Feng}
\affil{Key Laboratory for Particle Astrophysics, Institute of High Energy Physics, Chinese Academy of Sciences, 19B Yuquan Road, Beijing 100049, People's Republic of China}
\affil{University of Chinese Academy of Sciences, Chinese Academy of Sciences, Beijing 100049, People's Republic of China}

\author{L. D. Kong}
\affil{Key Laboratory for Particle Astrophysics, Institute of High Energy Physics, Chinese Academy of Sciences, 19B Yuquan Road, Beijing 100049, People's Republic of China}
\affil{University of Chinese Academy of Sciences, Chinese Academy of Sciences, Beijing 100049, People's Republic of China}

\author{P. J. Wang}
\affil{Key Laboratory for Particle Astrophysics, Institute of High Energy Physics, Chinese Academy of Sciences, 19B Yuquan Road, Beijing 100049, People's Republic of China}
\affil{University of Chinese Academy of Sciences, Chinese Academy of Sciences, Beijing 100049, People's Republic of China}

\author{S. N. Zhang}
\affil{Key Laboratory for Particle Astrophysics, Institute of High Energy Physics, Chinese Academy of Sciences, 19B Yuquan Road, Beijing 100049, People's Republic of China}
\affil{University of Chinese Academy of Sciences, Chinese Academy of Sciences, Beijing 100049, People's Republic of China}

\author{H. X. Liu}
\affil{Key Laboratory for Particle Astrophysics, Institute of High Energy Physics, Chinese Academy of Sciences, 19B Yuquan Road, Beijing 100049, People's Republic of China}
\affil{University of Chinese Academy of Sciences, Chinese Academy of Sciences, Beijing 100049, People's Republic of China}

\author{Z. X. Yang}
\affil{Key Laboratory for Particle Astrophysics, Institute of High Energy Physics, Chinese Academy of Sciences, 19B Yuquan Road, Beijing 100049, People's Republic of China}
\affil{University of Chinese Academy of Sciences, Chinese Academy of Sciences, Beijing 100049, People's Republic of China}

\author{Y. Huang}
\affil{Key Laboratory for Particle Astrophysics, Institute of High Energy Physics, Chinese Academy of Sciences, 19B Yuquan Road, Beijing 100049, People's Republic of China}
\affil{University of Chinese Academy of Sciences, Chinese Academy of Sciences, Beijing 100049, People's Republic of China}

\author{L. Ji}
\affil{School of Physics and Astronomy, Sun Yat-Sen University, Zhuhai, 519082, China}

\author{S. M. Jia}
\affil{Key Laboratory for Particle Astrophysics, Institute of High Energy Physics, Chinese Academy of Sciences, 19B Yuquan Road, Beijing 100049, People's Republic of China}

\author{X. Ma}
\affil{Key Laboratory for Particle Astrophysics, Institute of High Energy Physics, Chinese Academy of Sciences, 19B Yuquan Road, Beijing 100049, People's Republic of China}

\author{W. Yu}
\affil{Key Laboratory for Particle Astrophysics, Institute of High Energy Physics, Chinese Academy of Sciences, 19B Yuquan Road, Beijing 100049, People's Republic of China}
\affil{University of Chinese Academy of Sciences, Chinese Academy of Sciences, Beijing 100049, People's Republic of China}

\author{H. S. Zhao}
\affil{Key Laboratory for Particle Astrophysics, Institute of High Energy Physics, Chinese Academy of Sciences, 19B Yuquan Road, Beijing 100049, People's Republic of China}

\author{J. Y. Nie}
\affil{Key Laboratory for Particle Astrophysics, Institute of High Energy Physics, Chinese Academy of Sciences, 19B Yuquan Road, Beijing 100049, People's Republic of China}

\author{Y. L. Tuo}
\affil{Key Laboratory for Particle Astrophysics, Institute of High Energy Physics, Chinese Academy of Sciences, 19B Yuquan Road, Beijing 100049, People's Republic of China}

\author{S. Zhang}
\affil{Key Laboratory for Particle Astrophysics, Institute of High Energy Physics, Chinese Academy of Sciences, 19B Yuquan Road, Beijing 100049, People's Republic of China}

\author{J. L. Qu}
\affil{Key Laboratory for Particle Astrophysics, Institute of High Energy Physics, Chinese Academy of Sciences, 19B Yuquan Road, Beijing 100049, People's Republic of China}

\author{B. B. Wu}
\affil{Key Laboratory for Particle Astrophysics, Institute of High Energy Physics, Chinese Academy of Sciences, 19B Yuquan Road, Beijing 100049, People's Republic of China}


\begin{abstract}

We execute a detailed spectral-timing study of Cygnus X-1 in the low/hard, intermediate and high/soft states with the Hard X-ray Modulation Telescope observations. The broad band energy spectra fit well with the "truncated disk model" with the inner boundary of the accretion disk stays within $\sim$10 \textit R$_{\rm g}$ and moves inward as the source softens. Through studying of PDS, rms and Fourier-frequency component resolved spectroscopy, we find that the X-ray variations are generated in two different regions for each state. We discover that the major role that contributes to the X-ray variation is the hot corona rather than the accretion disk. We suggest a scenario with different corona geometry for each state based on the truncated disk geometry in which the corona wraps up the disk to form a sandwich geometry in the low/hard state, and then gradually moves away from the disk in direction that is perpendicular to the disk until forming a jet-like geometry in the high/soft state.

\end{abstract}


\section{Introduction} \label{Intro}

In general, X-ray binaries (XRBs) can exhibit Low/Hard state (LHS), Intermediate state (IMS) and High/Soft state (HSS) due to the variation in thermal and non-thermal emission. Each state shows distinct spectral and timing properties in X-rays and reflects accretion/ejection physics inherent to the accreting central objects. The origins of the X-rays as well as the variation of XRBs remain a matter of debate. 

A general consensus about the X-ray origin is that the soft X-ray is generated in an optically thick and geometrically thin accretion disk, and the hard X-ray is from the optically thin, large scale height corona. There exist many diverse models concerning the geometry in the vicinity of the central object. The extensively used model is the "truncated disk model" (\citealt{Ichimaru1977}; \citealt{Done2007}). The disk of the Shakura-Sunyaev type (\citealt{Shakura1973}) truncates at a radius, inside which is the hot corona. The hard X-rays are produced due to inverse Comptonization of soft disk photons by the hot corona in a power-law form. As the source evolves from hard to soft state, the mass accretion rate increases and the truncation radius moves inwardly, causing the geometry of the system changes as well as the interaction of the disk and corona.

The state transition of XRBs is probably a breakthrough point to the geometry of the inner region. We consider this issue from two aspects. On one hand, the broad band X-ray spectra reflect the composition and interactions of the emitters inside the system. On the other hand, in the context of the propagating mass accretion rate fluctuations model (\citealt{Ingram2016}), the X-ray variability in frequency domain could indicate the relative spatial distances of the emitters. Comparison and understanding of the key characteristics between different states may give clues concerning the geometry problem and the physical processes behind. Cygnus X-1 (Cyg X-1), a widely studied system, has the advantage of a large number of studies on individual spectral states, thus is well suited for studying and understanding transitions between different states and the accretion/ejection processes for accreting XRBs.

Cyg X-1 is a persistent bright high-mass X-ray binary with a primary object as confirmed BH (\citealt{Bowyer1965,Bolton1972,Webster1972}). The BH mass, the inclination and the distance of the system was estimated to be $21.2\pm2.2$ \textit M$_\odot$, $27.51\degree^{+0.77}_{-0.57}$, and $2.22^{+0.18}_{-0.17}$ kpc using VLBA observations (\citealt{Miller-Jones2021}). The dimensionless spin parameter of the BH \textit a$_*$ was estimated as $\sim$ 0.998 (\citealt{Zhao2021,Kushwaha2021}).

The energy spectrum of Cyg X-1 consists of direct component and reflected component. The direct component includes a thermal disk emission and a hard Compton tail from a hot corona. The reflection is due to hard X-rays irradiating the inner part of the optically thick accretion disk, and is dominated by a fluorescent Fe line at about 6--7 keV (depending on the ionization state of the disk) and a characteristic high-energy continuum peaking at about 20--30 keV (\citealt{George1991}). Cyg X-1 exhibits one of the best established examples of relativistic reflection study, especially in the soft state (\citealt{Duro2011,Fabian2012}). Many studies on X-ray reflection of Cyg X-1 have constrained the characteristic properties of the system such as the Fe abundance of the disk, the inner disk radius or the spin of the BH (\citealt{Zhao2021, Walton2016, Tomsick2014, Basak2017}). 

Power density spectrum (PDS) and fractional rms are the characteristics of X-ray variability. The PDS of Cyg X-1 is usually a broad band continuum of aperiodic variations without narrow features (\citealt{Rapisarda2017}). However, a recent work revealed the presence of a short-lived narrow quasi-periodic oscillation (QPO) in the soft state of Cyg X-1(\citealt{Yan2021}). The general shape of the PDS in the hard state of Cyg X-1 can be explained well as the sum of multiple broad Lorentzians (\citealt{Nowak2000}). Using the technique of Fourier-frequency component (FFC) resolved spectroscopy, \citet{Axelsson2018} decomposed the components of variability in the hard state of Cyg X-1 and found evidence of the existence of multiple different Comptonization regions. The soft state of Cyg X-1 does not show low-amplitude rapid variability, which is different from the soft state of most other XRBs (\citealt{Grinberg2014,Rapisarda2017}).

Most previous works on Cgy X--1 concentrated on spectral or timing analysis of one or two states. An overall spectral-timing study covering all typical states would shine light on the understanding of the X-ray characteristics and the accretion physics behind. The Hard X-ray Modulation Telescope (\citealt{ZhangSN2020,ZhangS2014}), also dubbed as \textit{Insight}{\rm -HXMT}, observed Cyg X-1 many times in pointing mode since its launching on June 15 2017. \textit{Insight}{\rm -HXMT} is designed with a wide energy coverage (1--250 keV) and a large effective area, especially at hard X-ray range (20--250 keV, 5100 cm$^2$), thus the observations of Cyg X-1 provide us a good opportunity with unique advantages in estimating the detailed spectral and timing properties of the source. In this paper, we conduct the comprehensive analysis of Cyg X-1 including the broad band energy spectra, PDS, rms spectra and FFC resolved spectra of three typical spectral states.

Our paper is organized as follows. In Section \ref{DATA} we summarize the data and analysis technique and then present our results in Section \ref{result}. We discuss our results in Section \ref{dis} and present the conclusions in Section \ref{conclu}.

\section{OBSERVATIONS AND DATA ANALYSIS} \label{DATA}

\textit{Insight}{\rm -HXMT}, which is the first Chinese X-ray astronomy satellite, has three kinds of main scientific payloads: the High Energy X-ray telescope (HE, 20--250 keV, 5100 cm$^2$, \citealt{Liu2020}) , the Medium Energy X-ray telescope (ME, 5--30 keV, 952 cm$^2$, \citealt{Cao2020}) and the Low Energy X-ray telescope (LE, 1--15 keV, 384 cm$^2$, \citealt{Chen2020}), capable of observations down to a time resolution of 1 ms, 276 $\mu \rm s$, and 25 $\mu \rm s$, respectively. 

We use the standard \textit{Insight}{\rm -HXMT} Data Analysis software (HXMTDAS) v2.04 to analyze the data. The data are filtered using the GTI recommended by the \textit{Insight}{\rm -HXMT} team; the elevation angle (ELV) is larger than 10 degree; the geometric cutoff rigidity (COR) is larger than 8 degree; the offset for the point position is smaller than 0.04 degree; data are used at least 300 s before and after the South Atlantic Anomaly (SAA) passage. Only the small field of view (FoV) mode of LE and ME is used, for preventing from the contamination of near-by sources and the bright earth. 

We also use the daily light curves from the Monitor of All-sky X-ray Image (\textit{MAXI}) mission (\citealt{Matsuoka2009}) to provide as a comparison and supplement\footnote{\textit{MAXI} observation of Cyg X-1 is in \url{http://maxi.riken.jp/star_data/J1958+352/J1958+352.html}.} of the \textit{Insight}{\rm -HXMT} observation. \textit{MAXI} is an all-sky X-ray monitor on the International Space Station, and started observations in August 2009. \textit{MAXI} has two types of X-ray slit cameras with wide FOVs and two kinds of X-ray detectors consisting of gas proportional counters covering the energy range of 2--30 keV and X-ray CCDs covering the energy range of 0.5--12 keV. 

\section{RESULTS} \label{result}

\subsection{Light Curves and Hardnesses}

\begin{figure*}
\centering
\includegraphics[width=0.7\textwidth]{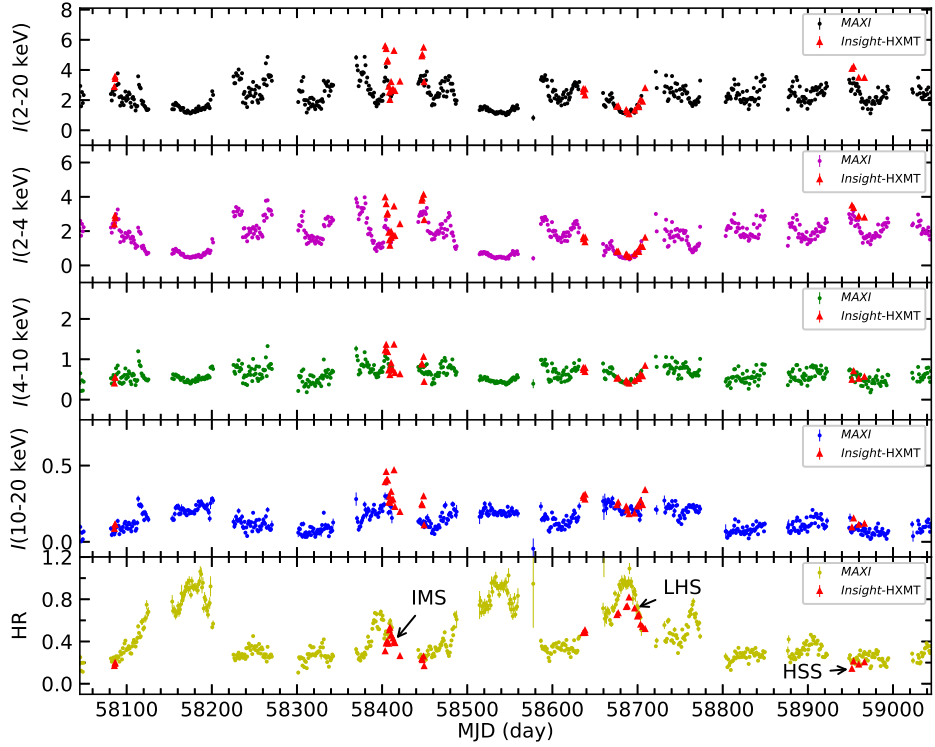}
\caption{From top to bottom, fluxes of energy ranges of 2--20 keV, 2--4 keV, 4--10 keV and 10--20 keV in units of photons cm$^{-2}$ s$^{-1}$, respectively, and HR corresponding to fluxes ratio of energy ranges of 4--10 keV to 2--4 keV. The three representative states analyzed are marked with arrows in the bottom panel.}
\label{fig:HMXTinMAXI}
\end{figure*}

We chose 40 \hxmt{} observation IDs (Obs. IDs; Table \ref{tab:40data}) from October 2018 to April 2020 that have simultaneous observations by \textit{MAXI}. Three representative states LHS (P020101217101, P020101217102), IMS (P010131502901) and HSS (P020101218601) are marked as bold in Table \ref{tab:40data} and with arrows in Figure \ref{fig:HMXTinMAXI}, among which the LHS is consist of two Obs. IDs that are obtained within 6 hours with similar fluxes and almost identical spectral shape in order to obtain good precision in timing analysis. Thus the figures in spectral analysis have 39 data points.

\begin{figure}
\centering
\includegraphics[width=0.35\textwidth]{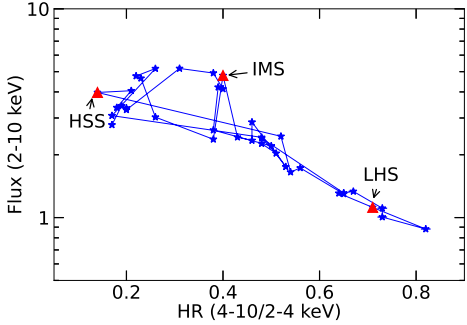}
\caption{HID of the 39 data points. The three representative states analyzed are marked with red triangles.}
\label{fig:HID}
\end{figure}

Figure \ref{fig:HMXTinMAXI} shows the photon fluxes of energy bands of 2--20 keV, 2--4 keV, 4--10 keV and 10--20 keV of Cyg X-1 from \textit{Insight}{\rm -HXMT} data in units of photons cm$^{-2}$ s$^{-1}$ which are compared to those from \textit{MAXI} data. For \hxmt{} data, the photon fluxes are obtained using the \textit {cpflux} method in spectral fitting with model constant$\times$\textit{TBabs}$\times$(\textit{diskbb}+\textit{cutoffpl}) (see Section \ref{Specfittext} for details) and energy ranges of 2.0--7.0 keV (LE), 10.0--20.0 keV (ME) and 28.00--100.0 keV (HE). The values of hardness ratio (HR) that are corresponding to flux ratio of energy ranges of 4--10 keV to 2--4 keV are also plotted. Each triangle of \textit{Insight}{\rm -HXMT} data represents one Obs. ID with duration of several hours, and each point of \textit{MAXI} data represents one day bin. Suggested by \citealt{Grinberg2013}, parameter $\Gamma$ (photon index) in spectral fitting with model constant$\times$\textit{TBabs}$\times$(\textit{diskbb}+\textit{relxill}) (see Section \ref{Specfittext} for details) is used to define states as low/hard ($\Gamma$ $<$ 2.0), intermediate ($\Gamma$ between 2.0 and 2.5) and high/soft ($\Gamma$ between 2.5 and 2.6), respectively. The hardness-intensity diagram (HID) of the 39 data points is shown in Figure \ref{fig:HID}.

\begin{table}
\caption{40 \hxmt{} Obs. IDs used in this work. MJD is the start time of the Obs. ID. HR is defined as the flux ratio of energy ranges of 4--10 keV to 2--4 keV. $\Gamma$ is the photon index in \textit{relxill}. State is defined according to $\Gamma$ as $<$ 2.0 (LHS), 2.0--2.5 (IMS) and 2.5--2.6 (HSS).}
\label{tab:40data}
\begin{tabular}{ccccc}
\hline\hline
Obs. ID.&MJD&HR&$\Gamma$&State\\\hline
P010131500704&58085.61&0.17&$2.49_{-0.05}^{+0.04}$&high/soft\\
P010131500802&58086.34&0.19&$2.57_{-0.01}^{+0.02}$&high/soft\\
P010131500803&58086.47&0.20&$2.56_{-0.02}^{+0.01}$&high/soft\\
P010131501901&58403.58&0.31&$2.03_{-0.01}^{+0.02}$&intermediate\\
P010131502001&58404.78&0.38&$2.00\pm{0.01}$&intermediate\\
P010131502101&58405.71&0.40&$1.974_{-0.009}^{+0.003}$&intermediate\\
P010131502201&58406.50&0.39&$1.979_{-0.003}^{+0.008}$&intermediate\\
P010131502301&58407.76&0.38&$1.95_{-0.01}^{+0.02}$&intermediate\\
P010131502401&58408.56&0.50&$1.83\pm{0.01}$&low/hard\\
P010131502501&58409.16&0.53&$1.77_{-0.02}^{+0.01}$&low/hard\\
P010131502601&58410.55&0.46&$1.91_{-0.02}^{+0.01}$&low/hard\\
P010131502602&58410.69&0.46&$1.89_{-0.02}^{+0.01}$&low/hard\\
P010131502801&58412.74&0.43&$1.91\pm0.02$&low/hard\\
\textbf{P010131502901}&58414.00&0.40&$2.00\pm{0.01}$&\textbf{intermediate}\\
P010131503001&58415.06&0.38&$1.96_{-0.01}^{+0.03}$&intermediate\\
P010131503302&58420.97&0.26&$2.07_{-0.01}^{+0.02}$&intermediate\\
P010131504701&58446.61&0.23&$2.25_{-0.01}^{+0.04}$&intermediate\\
P010131504801&58447.21&0.22&$2.25_{-0.02}^{+0.03}$&intermediate\\
P010131504901&58448.73&0.26&$2.25_{-0.01}^{+0.04}$&intermediate\\
P010131505001&58449.33&0.17&$2.31\pm0.02$&intermediate\\
P010131505801&58636.28&0.48&$1.88_{-0.02}^{+0.01}$&low/hard\\
P010131505802&58636.43&0.48&$1.89_{-0.02}^{+0.01}$&low/hard\\
P010131505901&58638.26&0.51&$1.83_{-0.03}^{+0.01}$&low/hard\\
P010131505902&58638.45&0.48&$1.87_{-0.04}^{+0.01}$&low/hard\\
P020101216002&58675.68&0.65&$1.64_{-0.06}^{+0.01}$&low/hard\\
P020101216101&58677.31&0.67&$1.62\pm{0.01}$&low/hard\\
P020101216601&58686.66&0.73&$1.58\pm{0.01}$&low/hard\\
P020101216701&58687.85&0.73&$1.58_{-0.07}^{+0.01}$&low/hard\\
P020101216801&58689.97&0.82&$1.58\pm0.01$&low/hard\\
\textbf{P020101217101}&58696.53&0.71&$1.61_{-0.02}^{+0.01}$&\textbf{low/hard}\\
\textbf{P020101217102}&58696.66&0.71&$1.62_{-0.03}^{+0.01}$&\textbf{low/hard}\\
P020101217301&58699.91&0.64&$1.68_{-0.03}^{+0.01}$&low/hard\\
P020101217401&58701.50&0.65&$1.67_{-0.02}^{+0.01}$&low/hard\\
P020101217501&58703.49&0.56&$1.75_{-0.04}^{+0.01}$&low/hard\\
P020101217601&58705.88&0.54&$1.77_{-0.02}^{+0.01}$&low/hard\\
P020101217701&58708.86&0.52&$1.83_{-0.01}^{+0.03}$&low/hard\\
\textbf{P020101218601}&58951.73&0.14&$2.61_{-0.02}^{+0.03}$&\textbf{high/soft}\\
P020101218701&58953.72&0.21&$2.57_{-0.01}^{+0.02}$&high/soft\\
P020101219001&58959.75&0.18&$2.52_{-0.02}^{+0.03}$&high/soft\\
P020101219301&58966.44&0.20&$2.54_{-0.02}^{+0.01}$&high/soft\\
\hline
\end{tabular}
\end{table}

\subsection{Broad Band Energy Spectra}

\subsubsection{Broad band spectral model}\label{Specfittext}

\textit{Insight}{\rm -HXMT} allows us to investigate spectra with a broad energy range. The energy bands adopted for spectral analysis are 2--10 keV (LE), 8--35 keV (ME) and 28--200 keV (HE), considering the calibration uncertainty and background level. The systematic error is set to 1\% (\citealt{Li2020}), and the errors of the parameters are calculated with Monte Carlo Markov Chain with a confidence interval of 90$\%$.

First, we fit the three representative spectra with model constant$\times$\textit{TBabs}$\times$(\textit{diskbb}+\textit{cutoffpl}) including a multiplicative constant to account for calibration uncertainties of the three instruments on board \textit{Insight}-HXMT, {\it TBabs} to account for the absorption due to the Inter Stellar Medium (ISM) (\citealt{Wilms2000}), a multi-temperature black body component {\it diskbb} to account for thermal emission from accretion disk (\citealt{Mitsuda1984}) and \textit{cutoffpl} to account for a power-law continuum with high-energy cutoff. The fit resulted in reduced $\chi^{2}$ ($\chi^{2}$ / degree of freedom) of 604.12/399 (LHS), 2477.65/399 (IMS) and 1147.1/392 (HSS) with residuals around 6--7 keV and $\sim$ 15--30 keV due to Iron line and reflection hump, respectively. Features below $\sim$ 3 keV were also noticed which could arise from the calibration uncertainties near the Si edge. We added a \textit{gaussian} with centroid energy fixed at 6.4 keV and $\sigma$ at $\sim$ 0.61 keV to improve the fit around 6--7 keV and received the reduced of $\chi^{2}$ 556.74/397 (LHS), 1648.07/397 (IMS) and 943.54/390 (HSS), respectively. Since the overall fit is still poor for IMS and HSS, we used the relativistic reflection model \textit{relxill} (\citealt{Garcia2014,Dauser2014}) that includes the direct primary power-law together with the reprocessing components arising from its interaction with the accretion disk to replace both the \textit{gaussian} and \textit{cutoffpl} components. Thus, the model used is constant $\ast$ {\it TBabs} $\ast$ ({\it diskbb} + {\it relxill}).

\begin{figure*}
\centering
\includegraphics[scale=0.2]{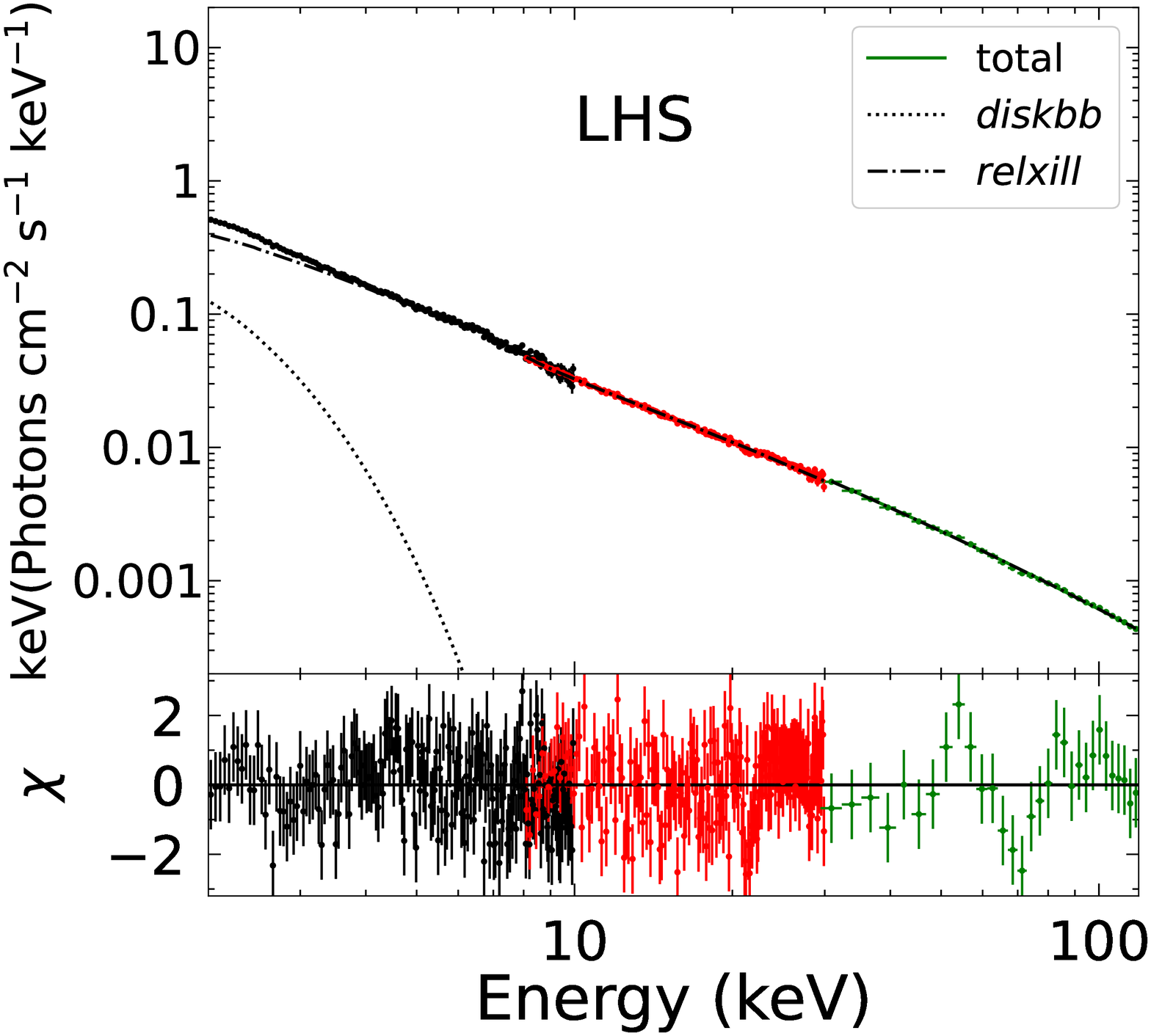}
\includegraphics[scale=0.2]{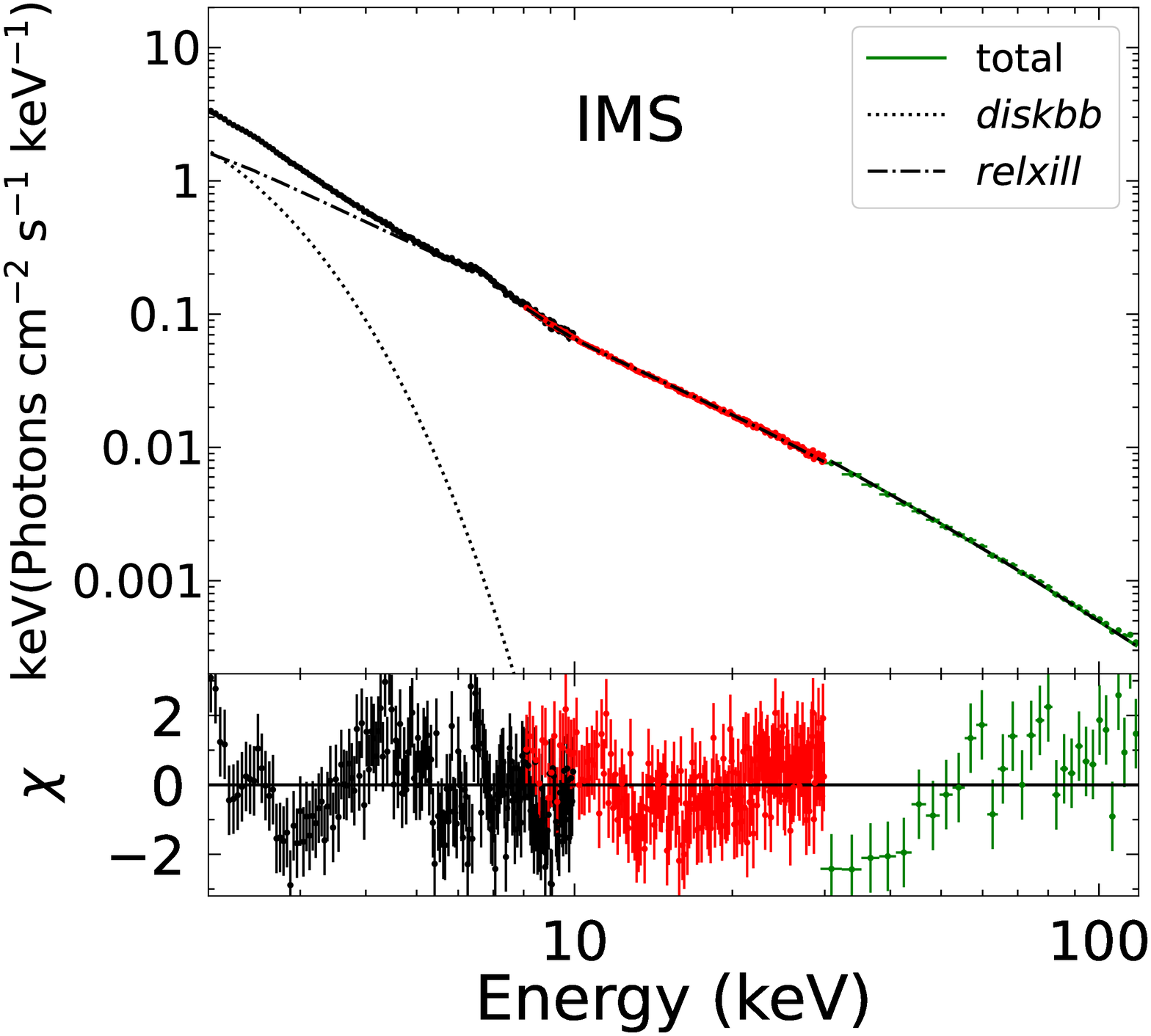}
\includegraphics[scale=0.2]{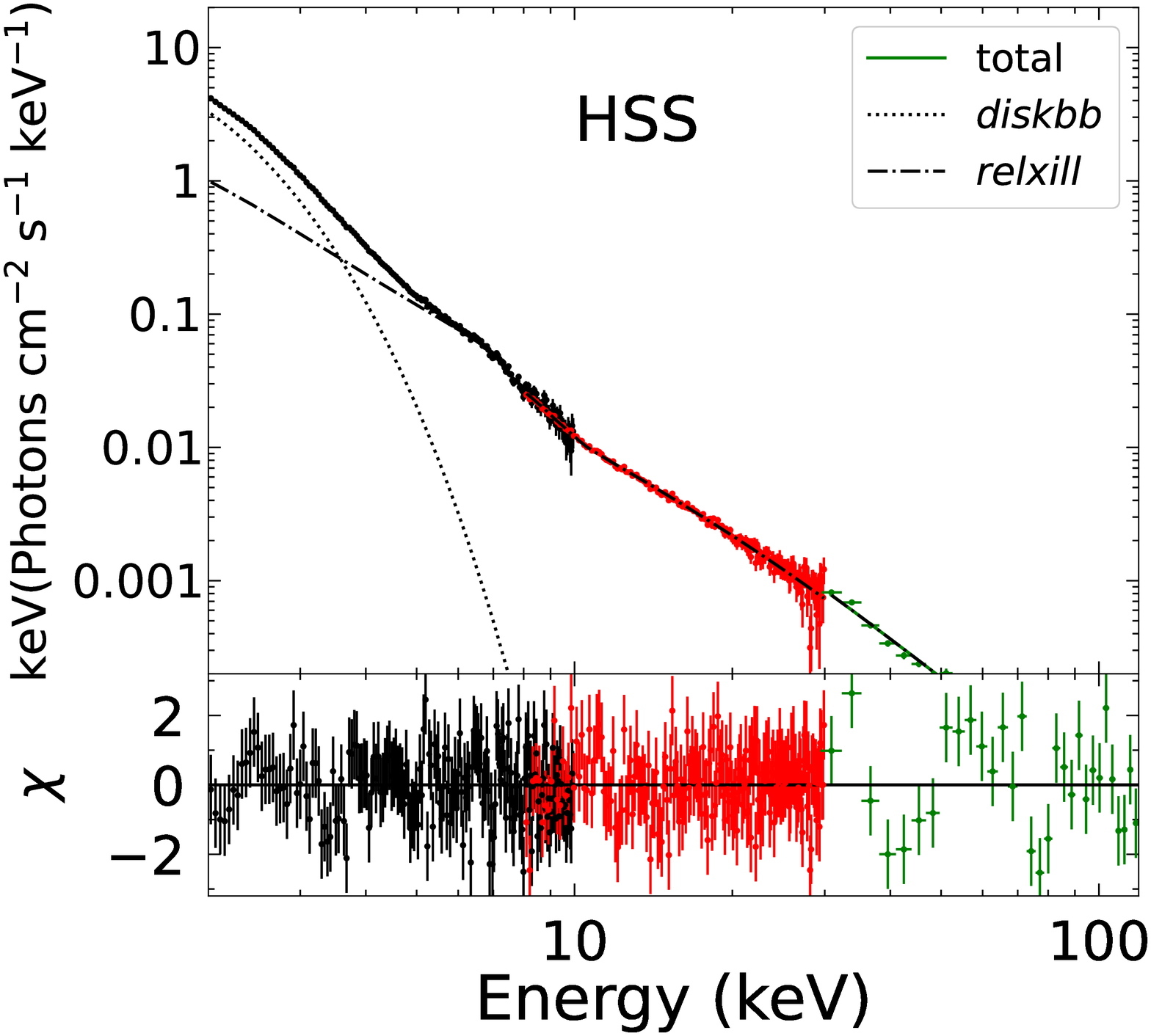}
\caption{From left to right, the broad band energy spectra of the three representative states using \textit{Insight}{\rm -HXMT} data. The black, red and green points with errors represent data form LE, ME and HE on-board \textit{Insight}{\rm -HXMT}, respectively.}
\label{fig:Spec}
\end{figure*}

The neutral absorption is fixed to $6.0\times10^{22}$ cm$^{-2}$ for simplicity (\citealt{Tomsick2014}). The spin parameter \textit a$_*$ is fixed to 0.998 which is the maximum value allowed in \textit {relxill} (\citealt{Zhao2021}). Radial emissivity Index1 and Index2 are both set to 3.0 and the radius \textit R$_{\rm br}$ at which the emissivity changes is set to equal the inner radius \textit R$_{\rm in}$. The outer boundary of the accretion disk \textit R$_{\rm out}$ is set to 400.0 in units of gravitational radius \textit R$_{\rm g}$. Redshift equals to zero. The high energy cutoff of the incident spectrum \textit E$_{\rm cut}$ is fixed to 300.0 keV. The iron abundance of Cyg X-1 is believed to be supersolar (\citealt{Hanke2009, Tomsick2014}). We found the parameter \textit A$_{\rm Fe}$ was hard to constrain in our fitting, thus we fixed it to 4.0 in units of solar abundance (\citealt{Parker2015, Walton2016, Basak2017, Tomsick2018}).

The reduced $\chi^{2}$ of all the 39 spectra were between 0.86 and 1.47. The two constants that indicate the calibration of ME and HE to LE (const$_{\rm ME}$, const$_{\rm HE}$) are found to be in the range of 0.88--0.99 and 0.84--1.12 respectively, which mean that the three instruments onboard \hxmt{} are well calibrated. Figure \ref{fig:Spec} shows the spectral fitting results of the three representative states with reduced $\chi^{2}$ of 1.07 (LHS), 1.31 (IMS) and 0.96 (HSS), respectively.

\begin{figure*}
\centering
\includegraphics[scale=0.4]{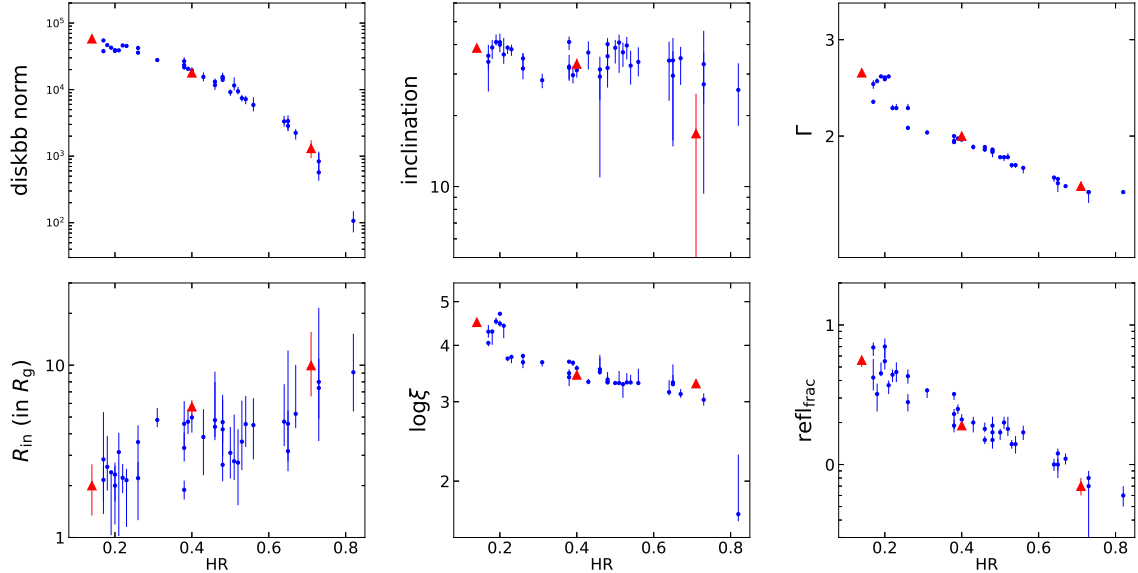}
\caption{From top to bottom and left to right, \textit{diskbb} normalization, inclination, $\Gamma$, \textit R$_{\rm in}$, log$\xi$ and reflection fraction of the 39 spectra. The red triangles in each panel indicate the three representative states.}
\label{fig:SpecPara}
\end{figure*}

Six free parameters, i.e. \textit{diskbb} normalization, inclination, $\Gamma$ (power-law index of the incident spectrum from the hot corona to the inner boundary of the disk), \textit R$_{\rm in}$ (inner boundary of the accretion disk in units of \textit R$_{\rm g}$), log$\xi$ (the ionization of the accretion disk at the inner edge) and refl$_{\rm frac}$ (reflection fraction that is defined as the fraction between photons emitted towards the disk and towards the observer) (\citealt{Dauser2016}), of all the 39 spectra are plotted in Figure \ref{fig:SpecPara}.

The inclination angle is well constrained to be 34.35$^{+0.79}_{-1.26}$ deg, indicating a moderate inclined disk. \textit R$_{\rm in}$ is located at a radius smaller than $\sim$ 10 times the \textit R$_{\rm g}$, and approaches to the central black hole as spectral softening. The trends of variations of the other three parameters with spectral softening are: (1) the incident spectrum becomes softer ($\Gamma$ decreases with increasing HR); (2) the ionization at the inner disk edge (log$\xi$) increases; (3) the reflection fraction (refl$_{\rm frac}$) increases.

\subsubsection{Evolution of the hardening factor and the effective disk temperature}

\begin{figure}
\centering
\includegraphics[scale=0.45]{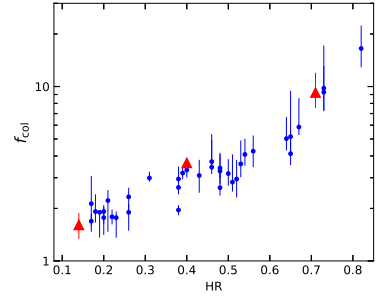}
\caption{\label{fig:hardeningfactor}Hardening factor as a function of HR. The red triangles in each panel indicate the three representative states.}
\end{figure}

Since the measured radiation spectrum of the accretion disk is affected and diluted by opacity and Comptonization, the effective temperature \textit T$_{\rm eff}$ of the disk can be corrected by \textit T$_{\rm eff}$ $=$ \textit T$_{\rm in}$ / \textit f$_{\rm col}$, where \textit f$_{\rm col}$ is the spectral hardening factor, and \textit T$_{\rm in}$ can be directly obtained from the spectral fitting resluts (\citealt{Shimura1995}). However, it is difficult to estimate \textit f$_{\rm col}$ when the power-law component is not negligible. Fortunately, the inner disk radius is also linked to \textit f$_{\rm col}$ by the expression \textit R$_{\rm in,norm}$ $=$ $\xi${~}\textit f$^{2}_{\rm col}$(\textit N / cos \textit i)$^{1/2}$\textit D$_{10}$, where \textit R$_{\rm in,norm}$ is the inner disk radius derived from the normalization of \textit {diskbb}, $\xi$ $=$ 0.412 is the geometry correction factor, \textit{N} is the \textit{diskbb} normalization (upper left panel in Figure \ref{fig:SpecPara}), \textit{i} is the inclination angle of the disk and \textit D$_{10}$ is the source distance in units of 10 kpc (\citealt{Mitsuda1984,Kubota1998}). Previous works have argued that for low mass XRBs {~}\textit f$_{\rm col}$ increases when the disk emission is relatively less dominant (\citealt{Merloni2000,Dunn2011}). It is plausible to assume a variable \textit f$_{\rm col}$ for Cyg X-1. Since we have known the fitted inner disk radius \textit R$_{\rm in}$ of \textit{relxill}, with the assumption that \textit R$_{\rm in}$ and \textit R$_{\rm in,norm}$ both equal the real inner disk radius, we obtain that \textit f$_{\rm col}$ increases with increasing HR as plotted in Figure \ref{fig:hardeningfactor}. The values of \textit f$_{\rm col}$ of Cyg X-1 are between 1.6 (the softest state) and 16.56 (the hardest state). Such large values in the intermediat and low/hard states are significantly larger than that in simulation (1.7-3; \citealt{Merloni2000}) and the low mass XRBs (1.6-2.6; \citealt{Dunn2011}) which are probably due to the complex accretion processes of Cyg X-1 as a high mass XRB.

\begin{figure}
\centering
\includegraphics[scale=0.45]{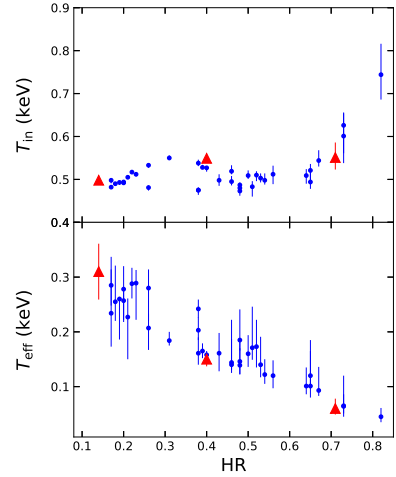}
\caption{\label{fig:Teff}Upper panel: Tin as a parameter of \textit{diskbb} in spectral fitting. Bottom panel: effective temperature of the disk emission. The red triangles in each panel indicate the three representative states.}
\end{figure}

Figure \ref{fig:Teff} shows \textit T$_{\rm in}$ (upper panel) and \textit T$_{\rm eff}$ (bottom panel) as functions of HR. This rough consideration of \textit f$_{\rm col}$ and \textit T$_{\rm eff}$ greatly improves the trend of variation of disk temperature to be consistent with the general consensus that the disk temperature rises as the system softens for most XRBs (\citealt{Kushwaha2021}).

\subsection{Power Density Spectra (PDSs)}\label{ChPDS}

The power density spectrum (PDS) is produced from 64 s data intervals with a time resolution of 1/125 s for a given energy range using HEASARC tool powerspec (XRONOS package), and is given in units of the squared fractional rms variability per frequency interval with Miyamoto normalization (\citealt{Miyamoto1991}) after subtracting the Possion noise. The frequency range we consider is from the minimum value of 1/64 Hz to the maximum value of 1/(2$\times$1/125) Hz.

\begin{figure*}[htbp]
\centering
\includegraphics[scale=0.2]{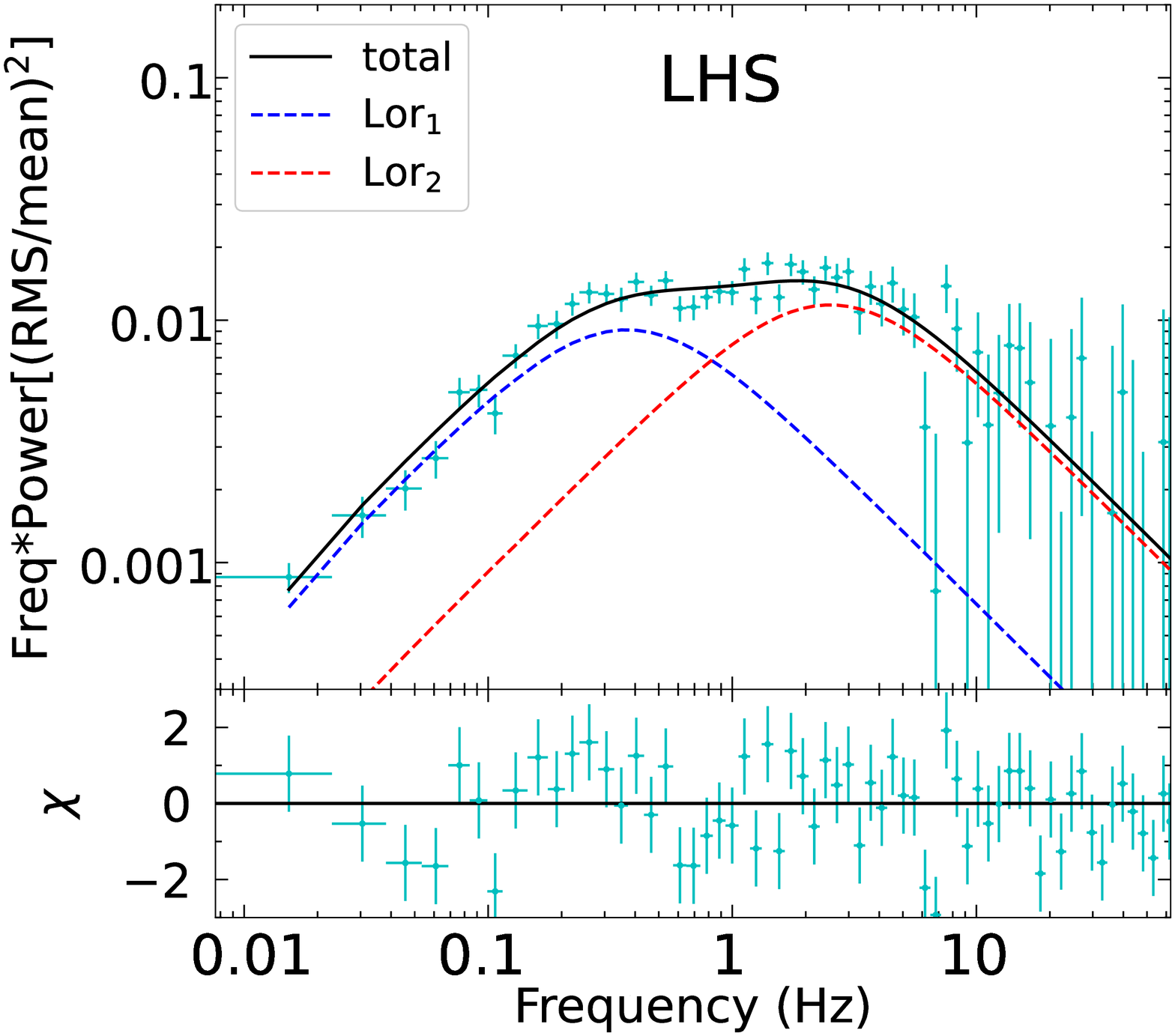}
\includegraphics[scale=0.2]{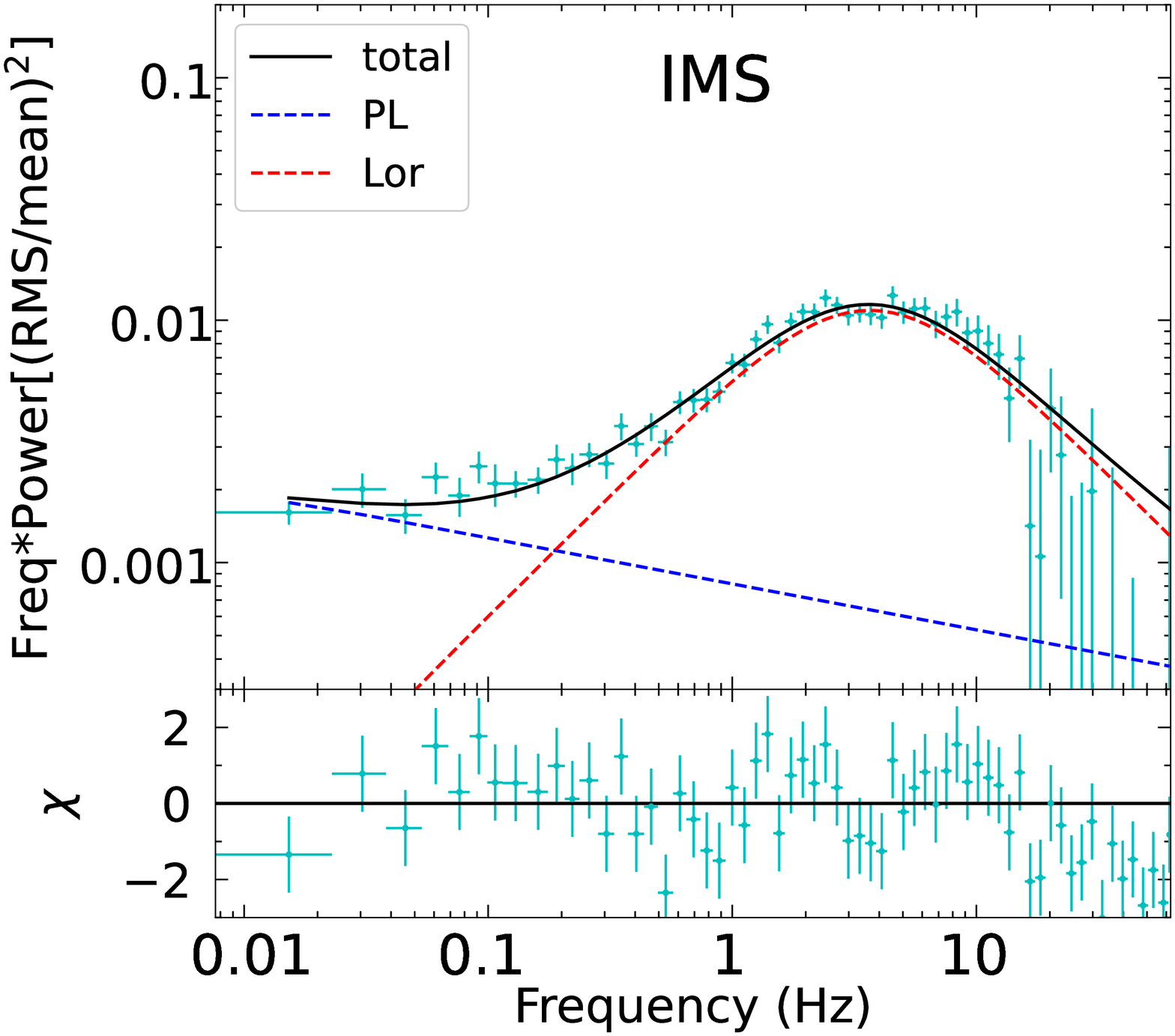}
\includegraphics[scale=0.2]{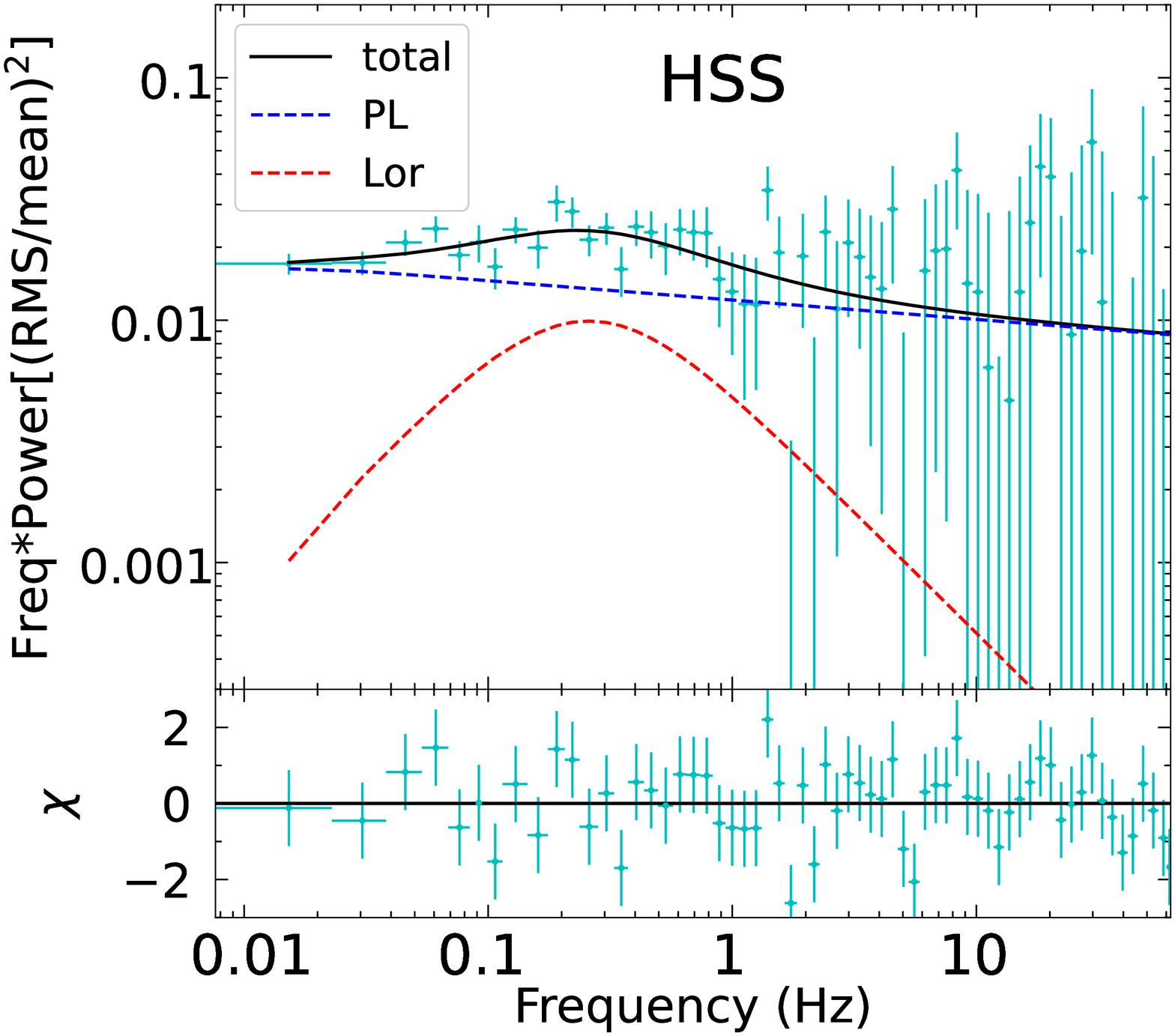}
\caption{\label{fig:PDS}From left to right, PDSs of the representative LHS, IMS and HSS using \textit{Insight}{\rm -HXMT}/ME data (6--38 keV) of Cyg X-1. The solid lines show the best fit and the colored dotted lines show the components.}
\end{figure*}

As indicated by previous works (\citealt{Nowak2000,Belloni2002}), the model applied to fit PDS of LHS is composed of two zero-centered Lorentzians. The Lorentzian profile is in the form of {$\propto$ 1 / [1 + (\textit f / \textit f$_0$)$^{2}$]}, where the centroid frequency is set to zero and half-width \textit f$_0$ represents the characteristic broken frequency of the feature. Taking PDS of 6--38 keV (ME) as an example, the two Lorentizians (Lor$_{1}$, Lor$_{2}$) of LHS in the left panel of Figure \ref{fig:PDS} are obviously characterized by different broken frequencies. Because the PDS of HSS is not well fitted by multi-Lorentzian model (\citealt{Pottschmidt2003}), we apply a Lorentzian (Lor) plus a power-law (PL) profile, among which the PL dominates in the whole frequency range, and the Lor is very weak and limited to lower broken frequency, as demonstrated in the right panel of Figure \ref{fig:PDS}. The PDS of IMS is also well fitted with a Lor plus a PL with the PL dominates in a narrower and lower frequency range, as demonstrated in the middle panel of Figure \ref{fig:PDS}. The models used for the PDSs of 6--38 keV (ME) and the best fit parameters are listed in Table \ref{tab:PDSmodel}. The systematic error is set to 0.0. The reduced $\chi^{2}$ are 1.30 (LHS), 1.60 (IMS) and 0.97 (HSS), respectively.

\begin{figure*}
\centering
\subfigure{
\begin{minipage}[b]{0.2\linewidth}
\includegraphics[angle=0, scale=0.18]{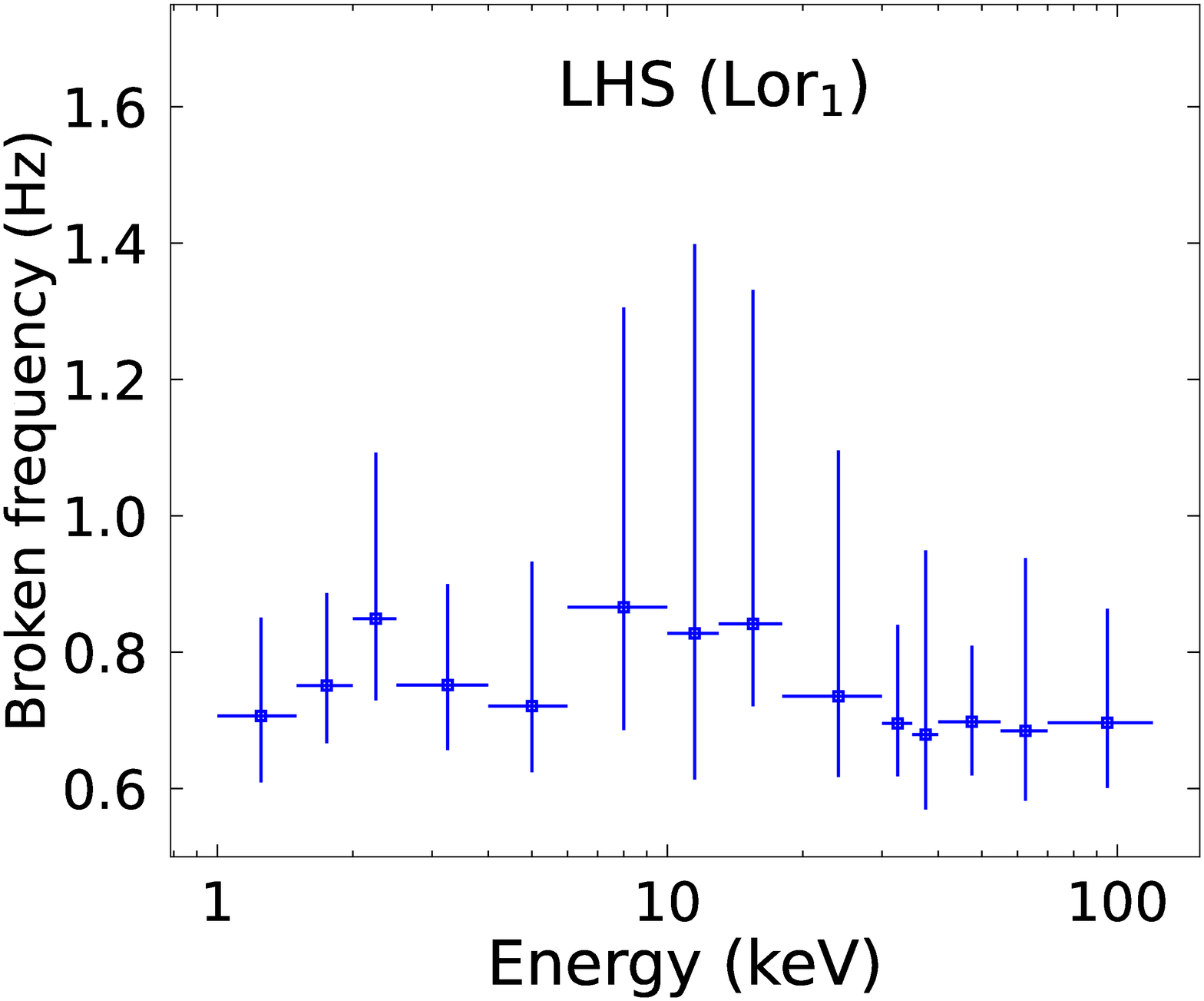}
\includegraphics[angle=0, scale=0.18]{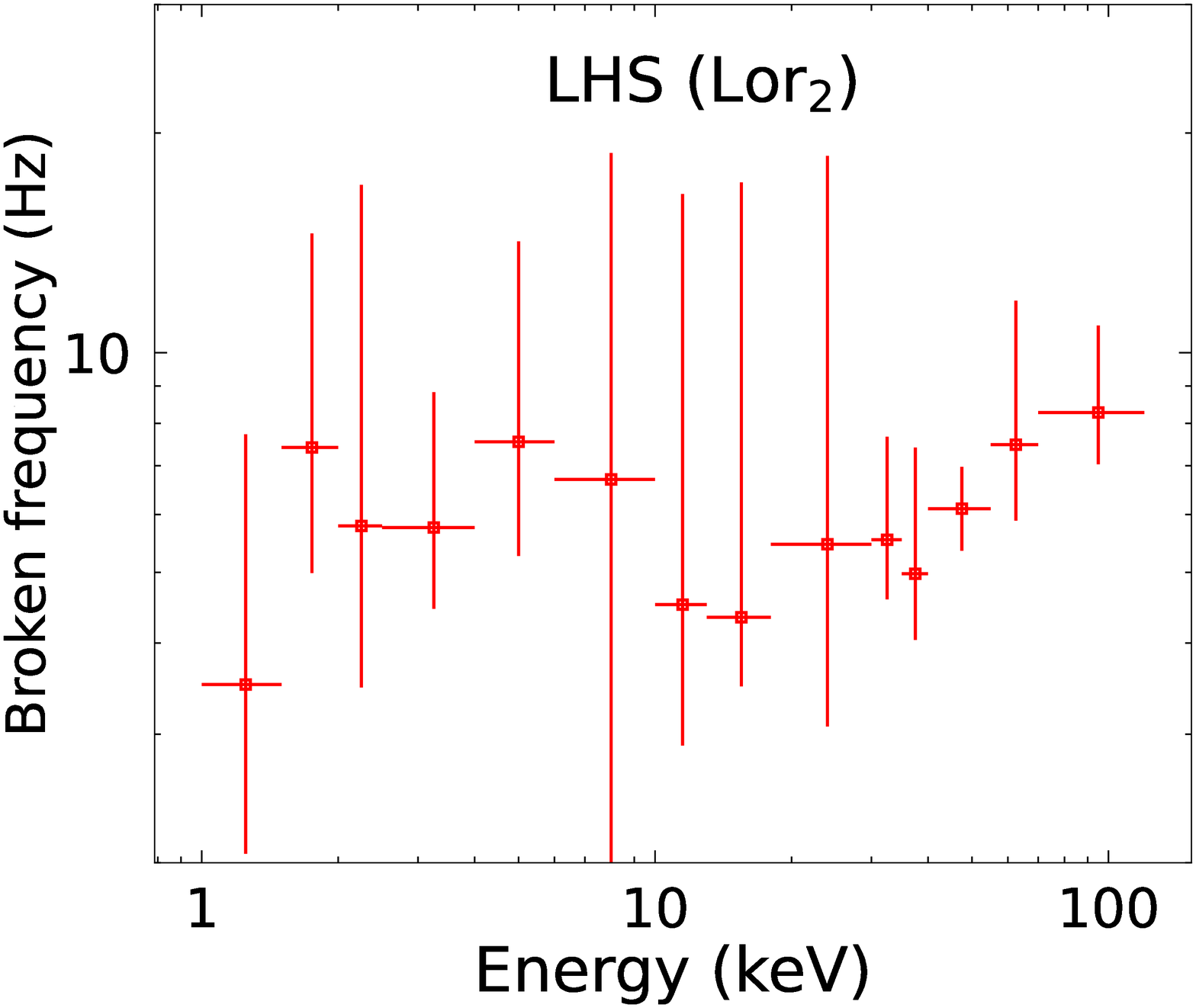}
\end{minipage}}
\hspace{15 mm}
\subfigure{
\begin{minipage}[b]{0.2\linewidth}
\includegraphics[angle=0, scale=0.18]{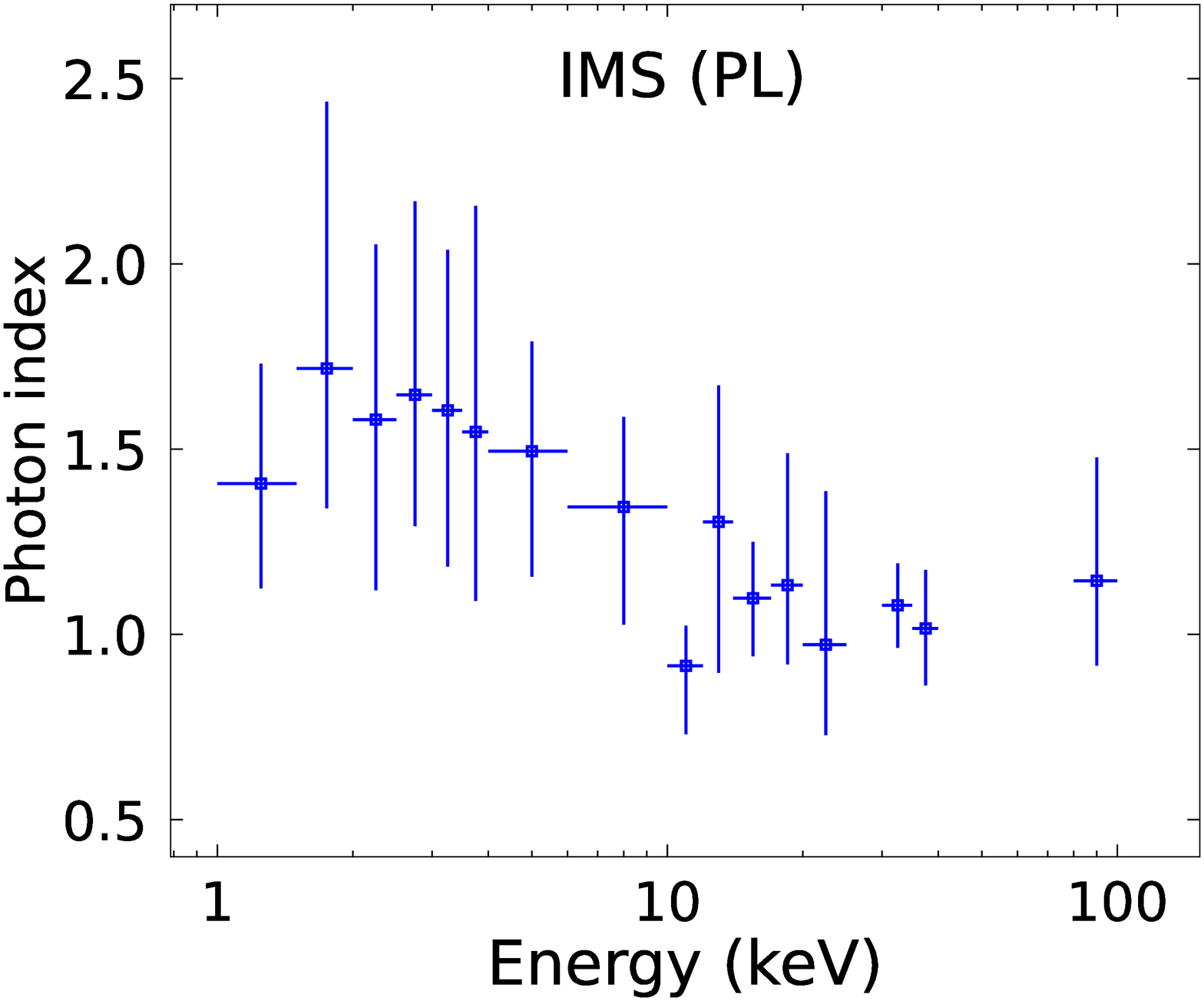}
\includegraphics[angle=0, scale=0.18]{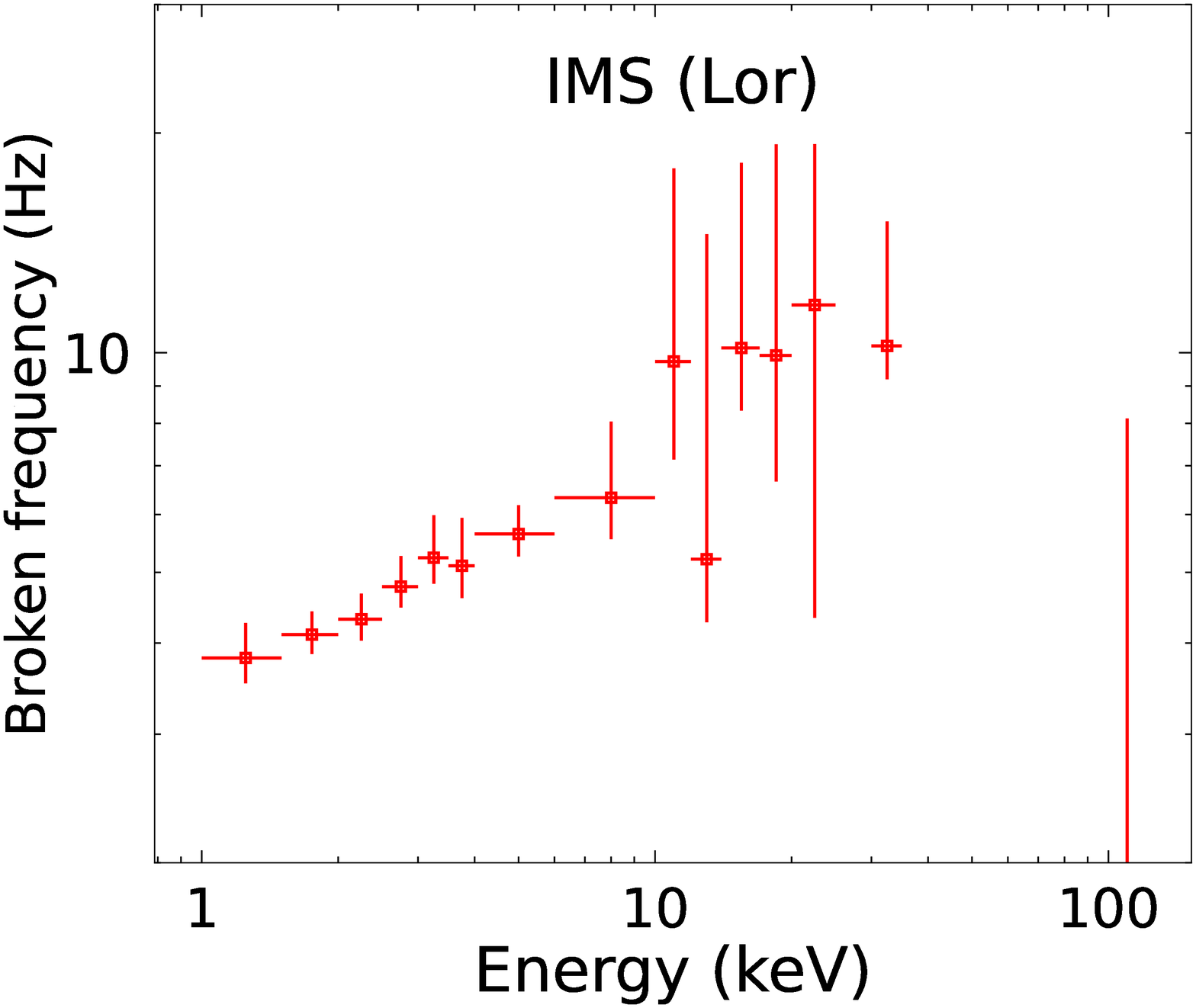}
\end{minipage}}
\hspace{15 mm}
\subfigure{
\begin{minipage}[b]{0.2\linewidth}
\includegraphics[angle=0, scale=0.18]{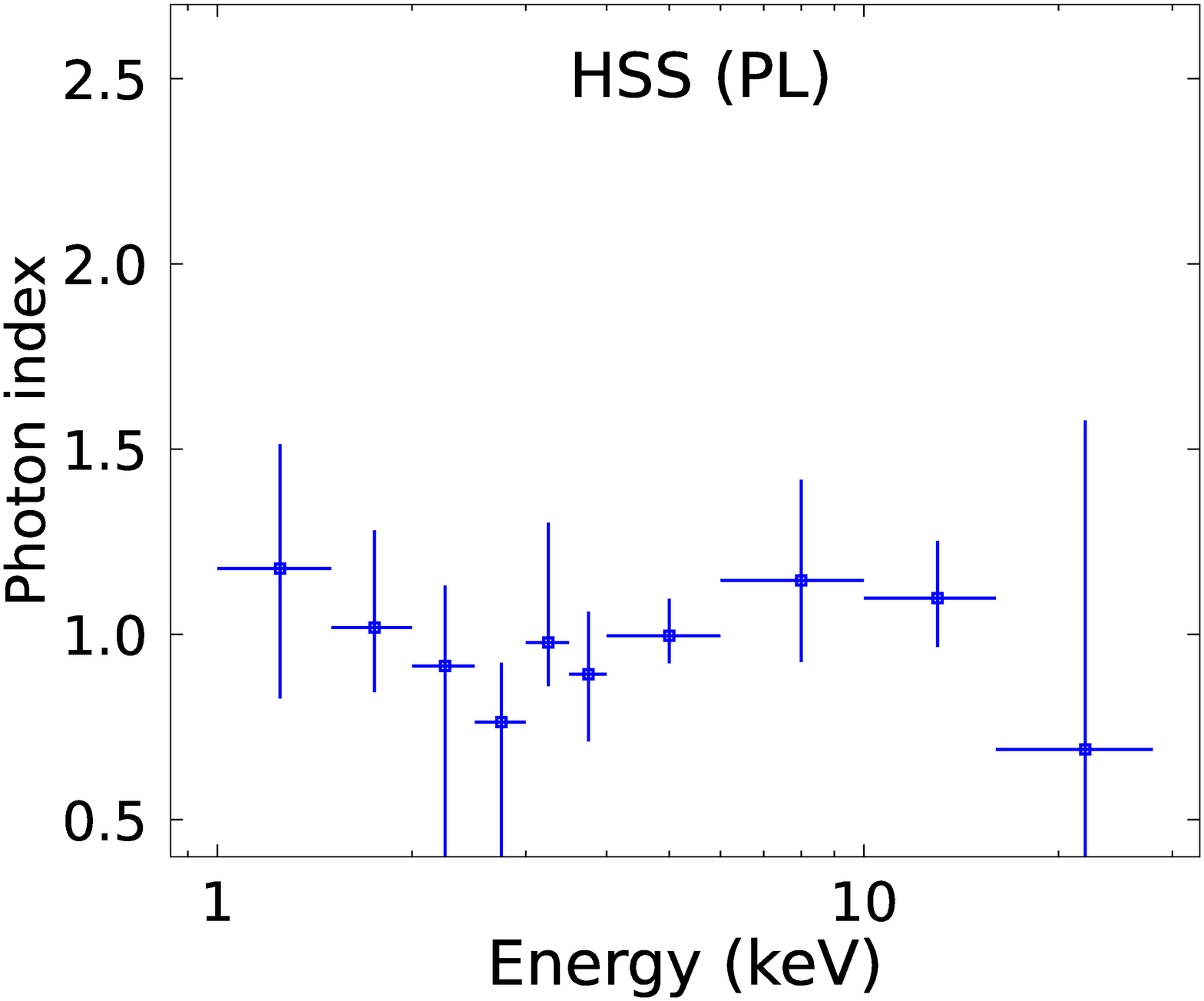}
\includegraphics[angle=0, scale=0.18]{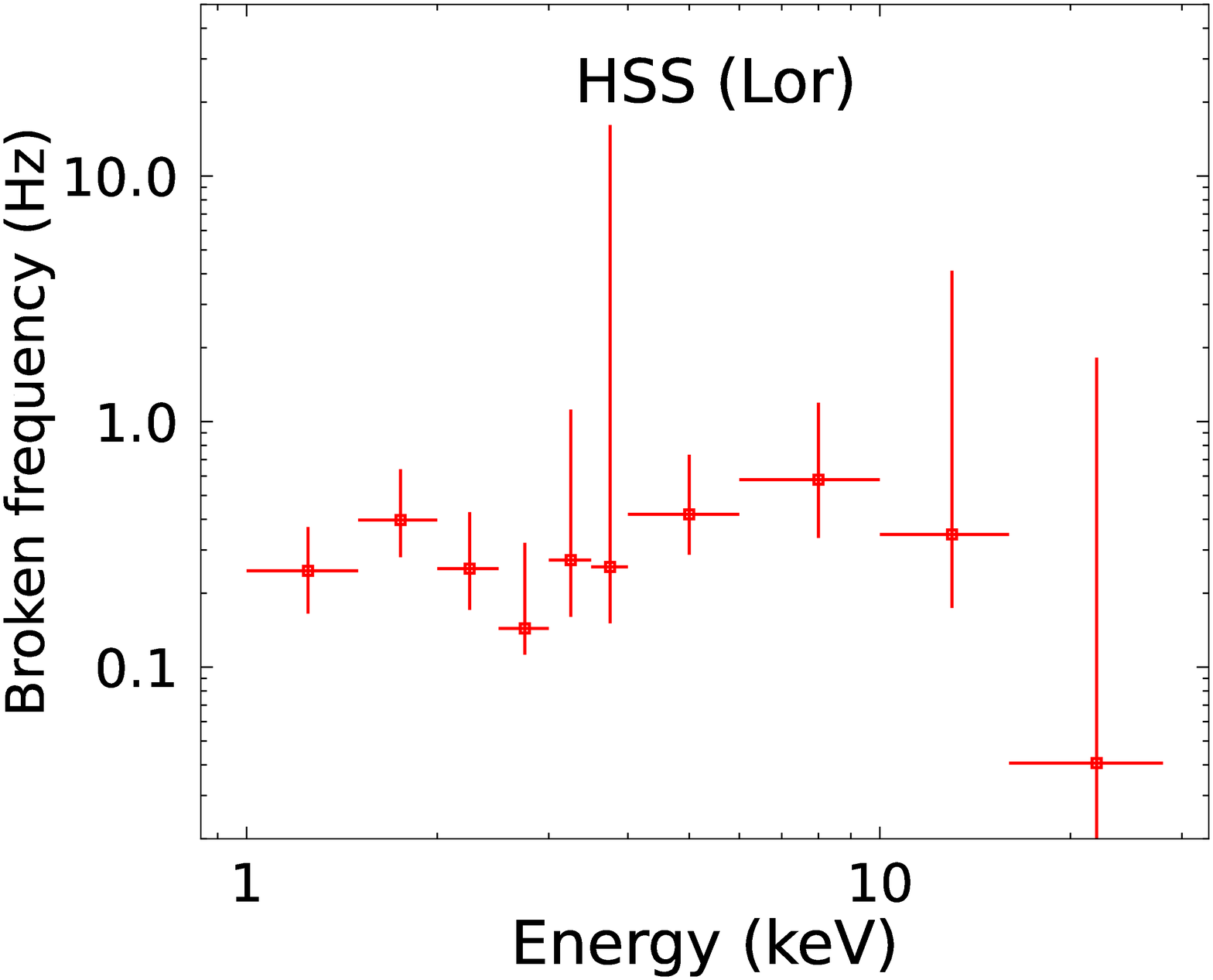}
\end{minipage}}
\caption{\label{fig:PDScompparfig}The broken frequencies of the two Lorentzians of the representative LHS (left); the photon index of the power-law profile and the broken frequency of the Lorentzian profile of the representative IMS (middle) and HSS (right).}
\end{figure*}

To investigate how the characteristics of the PDSs' components change with photon energy, we carefully divided the 1-120 keV energy range into small bins under consideration of data quality for each representative state. We then fit the PDS of each energy bin with the model mentioned in the above paragraph and plotted the characteristics of the PDSs' components as functions of energy which are shown in Figure \ref{fig:PDScompparfig}. These characteristics include the broken frequencies of the Lorentzians and the photon indexes of the power-laws. The broken frequencies of Lor$_1$ and Lor$_2$ in LHS hardly show dependence upon energy. Lor in IMS increases with increasing energy, with the correlation in lower energy range more prominent. The photon index of PL in IMS decreases with increasing energy. Both of the photon index of PL and broken frequency of Lor in HSS do not change with energy.

It is worth mentioning that the above PDSs of 6--38 keV are estimated by subtracting the white noise. However, in the following section we investigate the fractional rms of PDSs without the white noise subtraction. We then apply an additional power-law profile with photon index fixed to zero to fit the Poisson noise. 

\begin{table*}
\caption{\label{tab:PDSmodel}The best fit free parameters of PDSs of 6--38 keV (ME) of the representative LHS, IMS and HSS.}
\centering
\begin{tabular}{c c c c c}
 \\\hline \hline
Component&Parameter&LHS&IMS&HSS\\\hline
Lor$_{1}$ &Broken frequency [Hz]& $0.74^{+0.12}_{-0.08}$ & - & $0.51^{+0.36}_{-0.17}$\\
 & norm & $0.029^{+0.004}_{-0.003}$ & - & $0.031^{+0.014}_{-0.013}$\\\hline
Lor$_2$&Broken frequency [Hz]& $5.03^{+1.05}_{-0.73}$ & $7.31^{+0.47}_{-0.43}$ & -\\
 & norm & $0.036^{+0.003}_{-0.004}$ & $0.035\pm0.002$ & -\\\hline
PL &Photon Index&-& $1.19\pm0.08$ & $1.08^{+0.10}_{-0.06}$\\
 & norm & - & $8.17\times10^{-4+2.61\times10^{-4}}_{-2.15\times10^{-4}}$ & $0.012^{+0.003}_{-0.004}$ \\\hline \hline
$\chi_{\rm red}^{2}(d.o.f)$& &76.80/59&94.54/59&56.95/59\\\hline
\end{tabular}
\end{table*}

\subsection{Fractional Root Mean Square (rms)}\label{rms}

\begin{figure*}
\centering
\includegraphics[scale=0.18]{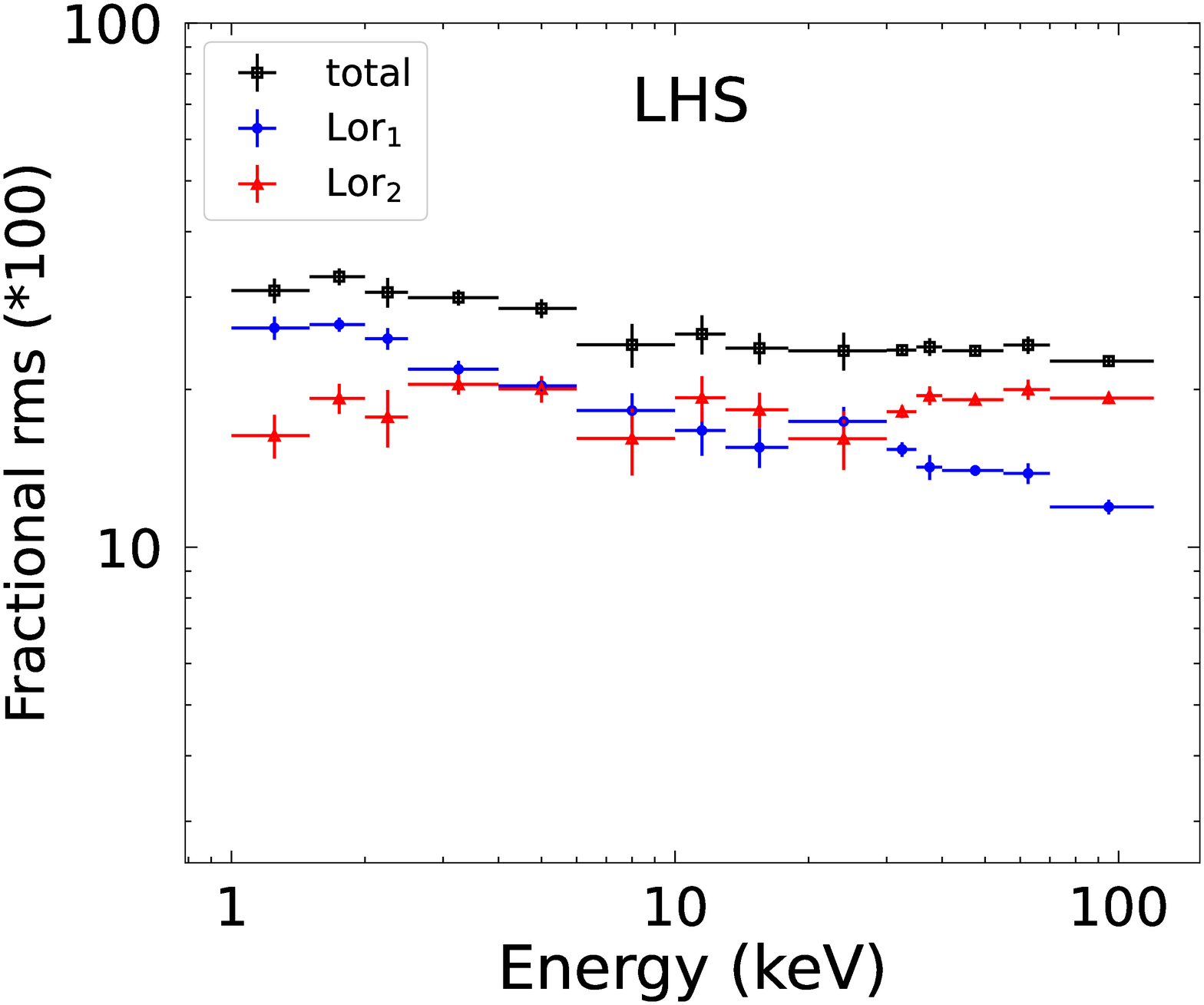}
\includegraphics[scale=0.18]{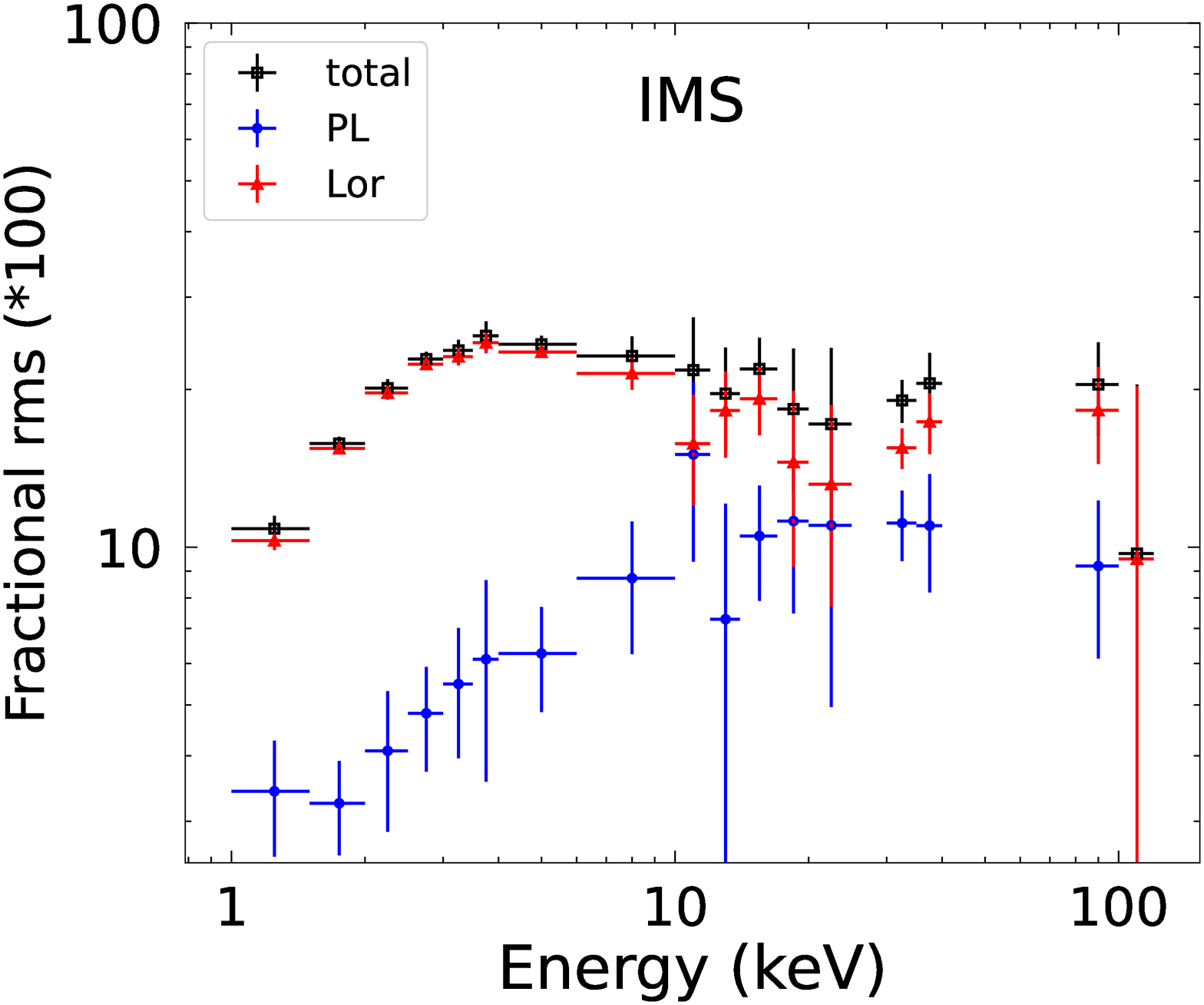}
\includegraphics[scale=0.18]{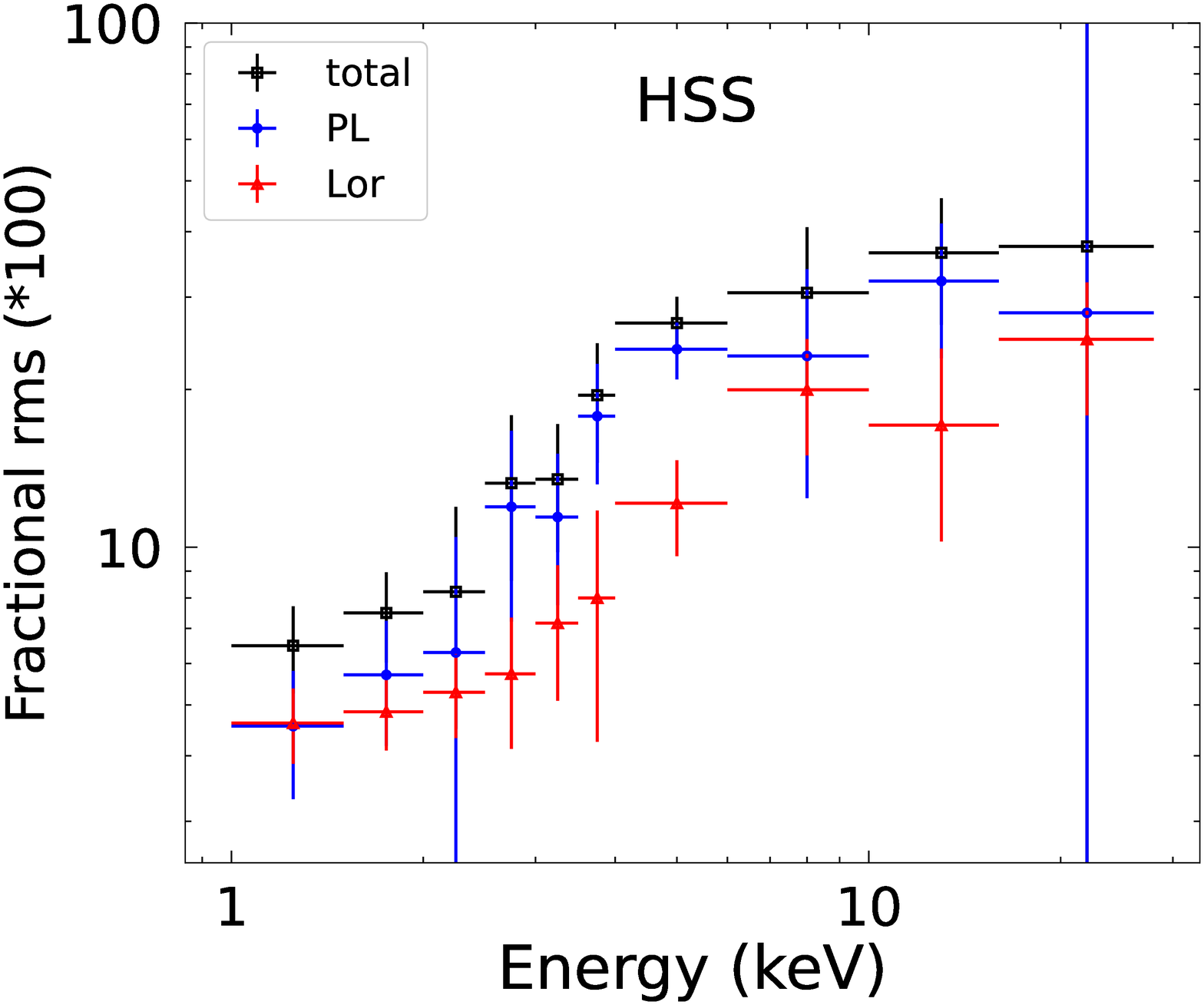}
\caption{\label{fig:RMSspec}The total fractional rms amplitude and the components of the three representative states.}
\end{figure*}

To quantitatively investigate the variation as a function of photon energy, we estimated the rms from the PDSs for each state. The rms of a single PDS component at a given energy range is obtained by integrating the power density over the frequency range considered and acquiring the square root, which is shown as colored points with errors in Figure \ref{fig:RMSspec}. The total rms of a spectral state is the square root of the summed integration of the two components in each energy range, which is shown as black points with errors. The frequency range considered in rms calculation covers the whole frequency range of the PDSs as mentioned in Section \ref{ChPDS}.

The total rms and the components behave differently for different spectral states. In LHS, the total rms slightly decreases with increasing energy with the maximum value 32.8\% in 1.5--2.0 keV and the minimum value 22.6\% in 70.0--120.0 keV. The rms amplitude of Lor$_2$ stays roughly constant. In IMS, the total rms rises from 10.8\% in 1.0--1.5 keV to the maximum value of 25.3\% in 3.5--4.0 keV and then stays roughly constant with the Lor as the dominator in the whole energy range rather than PL. In HSS, the rms could only be measured in a narrower energy range than in the other two states because the signal becomes too weak at energy higher than 28.0 keV. The total rms amplitude rises with energy monotonically from 6.5\% in 1.0--1.5 keV to 37.5\% in 16.0--28.0 keV with the PL as the main contributor in the whole energy range considered. The component Lor shows quite a large amplitude of $>$ 10$\%$ at 4.0--28.0 keV.

\subsection{FFC Resolved Spectra (RMS spectra)}\label{RMS}

The RMS spectrum can be regarded as the real energy spectrum of the variable part of the X-ray flux. Following the same technique as in \citet{Revnivtsev1999} and \citet{Axelsson2018}, we denoted it by the formula \textit{RMS(E) = rms(E) $\times$ R(E) / Area(E)}, where \textit{rms(E)} is the fractional rms as a function of energy, \textit{R(E)} is the corresponding counts rate and \textit{Area(E)} is the effective area of the corresponding instrument at that energy.

\begin{figure*}
\centering
\includegraphics[scale=0.20]{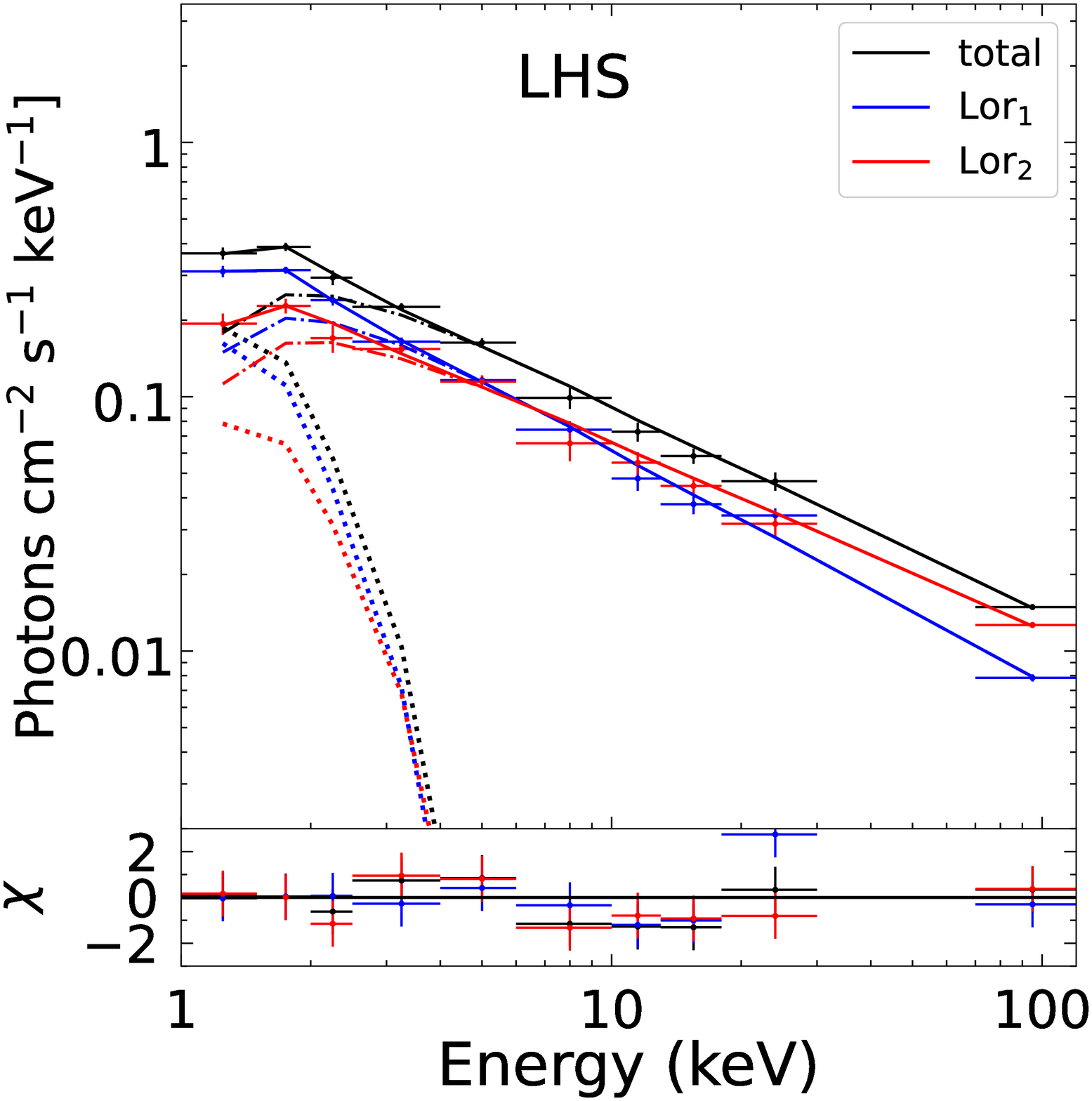}
\includegraphics[scale=0.20]{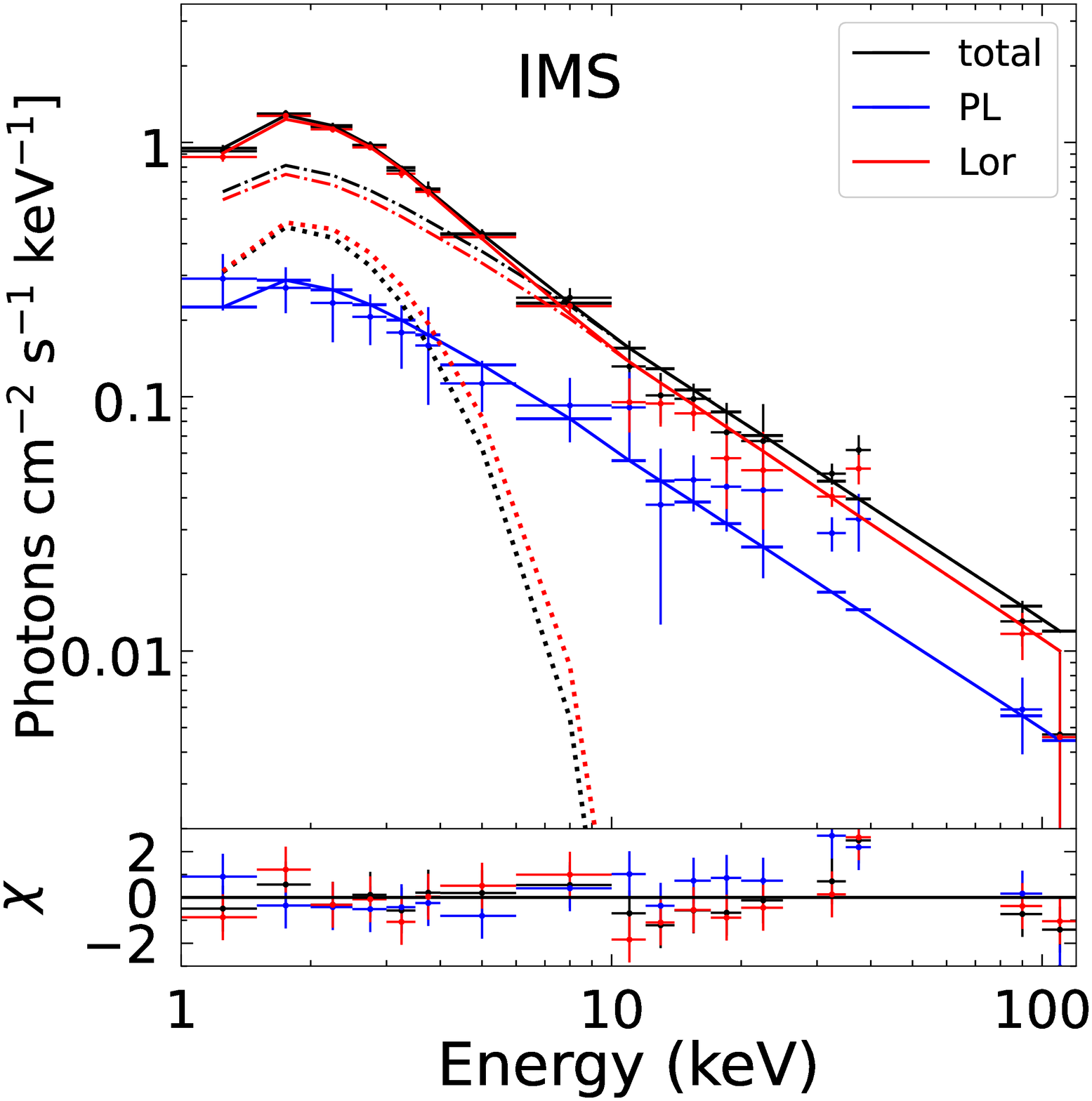}
\includegraphics[scale=0.20]{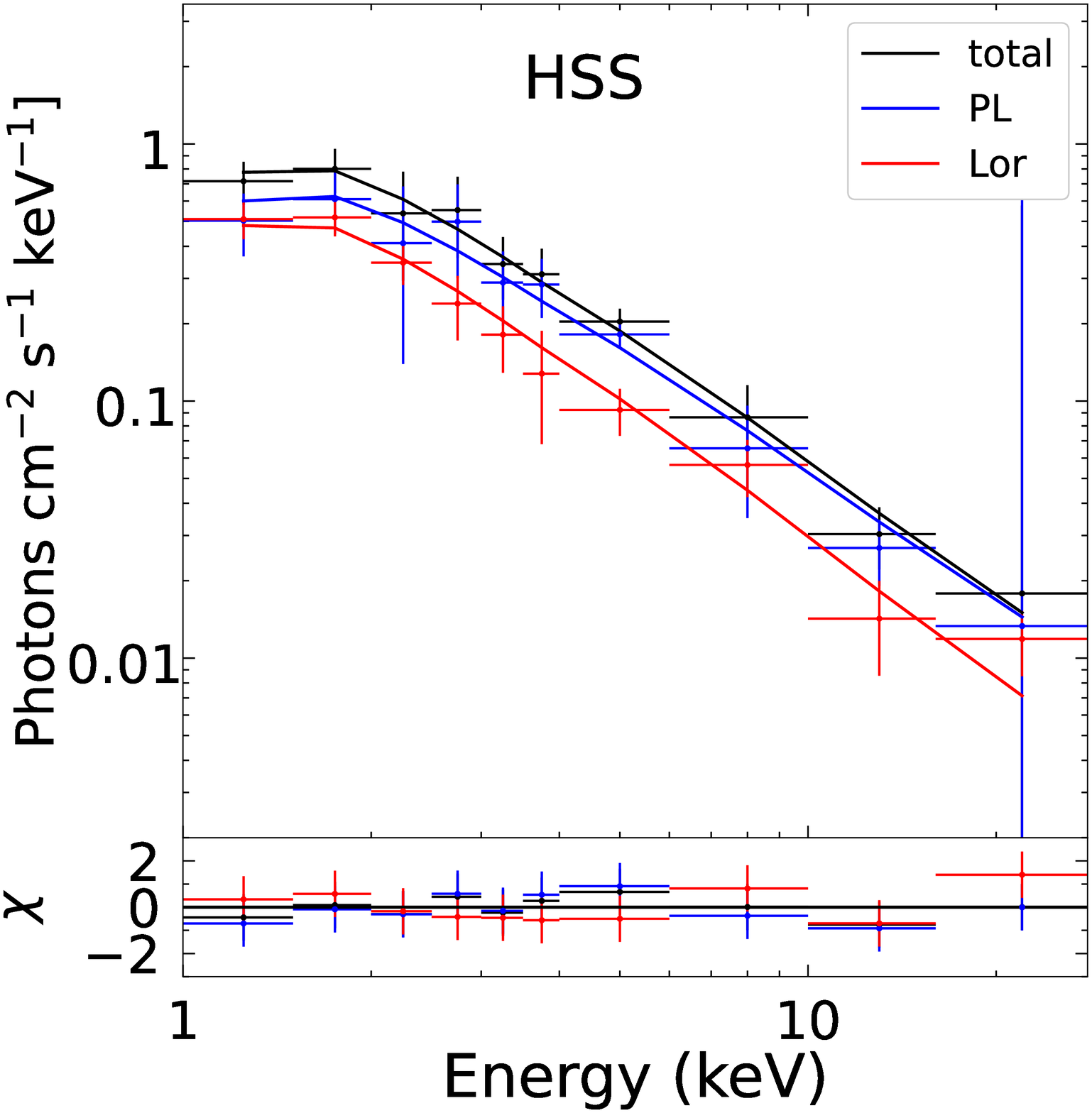}
\caption{\label{fig:RMSfit}Fitting diagrams of the RMS spectra of the three states. In each panel, the data points with errors are the total RMS spectrum of the state (black) and its two components (blue and red)). The black, blue and red lines are the best fit results. The dot-dashed lines and dotted lines with the corresponding colors are the power-law and \textit{diskbb} components.}
\end{figure*}

We fit the RMS spectrum with a single power-law function. If the reduced $\chi^{2}$ is higher than 2.0, a \textit{diskbb} component is added and \textit{F}-test is executed. The neutral absorption is also fixed to $6.0\times10^{22}$ cm$^{-2}$, and the systematic error is set to 0.0. The fitting diagrams of the total RMS spectra and the components are shown in Figure \ref{fig:RMSfit}. The models and the best fit parameters are listed in Table \ref{tab:RMSfitpara}. 

The fitting results show that a disk-like component is required to the X-ray variation in LHS and IMS. As the system turns softer, the disk-like component disappears, leaving only the power-law component contributing to the variability. This is consistent with the general scenario that in HSS the disk emission is usually stable, and most of the variability is associated with the hard spectral component (\citealt{Gilfanov2010} and reference herein). 

It is worth pointing out that the power-law component that contributes to the total RMS spectrum (column 3 of Table \ref{tab:RMSfitpara}) becomes softer as the system softens, which is consistent with the trend of the parameter $\Gamma$ in energy spectral fitting in Figure \ref{fig:SpecPara}, though the values are overall larger. This trend provides evidence that the fitting of the RMS spectra is plausible.

\begin{table*}
\small
\centering
\caption{\label{tab:RMSfitpara}Spectral fitting parameters of the RMS spectra of the 3 states.}
\begin{tabular}[c]{c c c c c c c}
\\\hline \hline
State&RMS spectrum&Component&Parameter&Value&$\chi_{\rm red}^{2}(d.o.f)$&\textit{F}-test (\textit{p}-value)\\\hline 
LHS&total&\textit{diskbb}&$T_{\rm in}$ [keV]& $0.33^{+0.03}_{-0.07}$ &6.52/6&$1.67\times10^{-4}$\\
  &   &   & norm &$9.84\times10^{3+3.22\times10^4}_{-4.08\times10^3}$&  & \\
  &   &power-law& Photon Index & $1.82^{+0.03}_{-0.02}$ &  & \\
  &   &   & norm &$0.59^{+0.06}_{-0.04}$ &  & \\
  \cline{2-7}
  & Lor$_1$ &\textit{diskbb}&$T_{\rm in}$ [keV]& $0.32^{+0.03}_{-0.05}$ & 10.47/6 & $3.96\times10^{-4}$\\
  &   &   & norm & $1.09\times10^{4+2.04\times10^4}_{-4.68\times10^3}$ &  & \\
  &   &power-law& Photon Index & $1.92^{+0.03}_{-0.02}$ &  & \\
  &   &   & norm & $0.51^{+0.05}_{-0.03}$ &  & \\
    \cline{2-7}
  & Lor$_2$ &\textit{diskbb}&$T_{\rm in}$ [keV]& $0.37^{+0.09}_{-0.13}$ & 6.92/6 & $1.04\times10^{-2}$\\
  &   &   & norm &$2.53\times10^{3+3.09\times10^4}_{-1.65\times10^3}$&  & \\ 
  &   &power-law& Photon Index & $1.75^{+0.03}_{-0.02}$ &  & \\
  &   &   & norm & $0.37^{+0.05}_{-0.03}$ &  & \\\hline \hline
IMS & total &\textit{diskbb}&$T_{\rm in}$ [keV]& $0.77\pm0.10$ & 13.30/13 & $1.46\times10^{-3}$\\
  &   &   & norm &$5.14\times10^{2+3.31\times10^2}_{-2.64\times10^2}$&  & \\
  &   & power-law & Photon Index & $2.12^{+0.09}_{-0.06}$ & - & -\\
  &   &   & norm & $2.25^{+0.61}_{-0.35}$ &  & \\
    \cline{2-7}
  & Lor &\textit{diskbb}&$T_{\rm in}$ [keV]& $0.83\pm0.07$ & 18.71/13 & $2.08\times10^{-4}$\\
  &   &   & norm &$4.04\times10^{2+1.65\times10^2}_{-1.63\times10^2}$&  & \\
  &   &power-law& Photon Index & $2.14^{+0.08}_{-0.06}$ &  & \\
  &   &   & norm & $2.10^{+0.46}_{-0.32}$ &  & \\
    \cline{2-7}
  & PL &power-law& Photon Index & $2.10^{+0.09}_{-0.07}$ & 26.86/15 & -\\
  &   &   & norm & $0.79^{+0.17}_{-0.14}$ &  &  \\\hline \hline
HSS & total &power-law& Photon Index & $2.75^{+0.18}_{-0.15}$ &1.63/8 & -\\
  &   &   & norm &$3.07^{+0.78}_{-0.65}$ &  & \\
    \cline{2-7}
  & Lor &power-law& Photon Index & $2.85^{+0.23}_{-0.15}$ & 4.54/8 & -\\
  &   &   & norm & $1.93^{+0.49}_{-0.36}$ &  & \\
    \cline{2-7}
  & PL &power-law& Photon Index & $2.67^{+0.20}_{-0.18}$ & 3.04/8 & -\\
  &   &   & norm & $2.34^{+0.73}_{-0.60}$ &  & \\\hline
\end{tabular}
\end{table*}

\section{DISCUSSION}\label{dis}

The origins of the variation and broad band noise in X-ray emission of XRBs is still a matter of debate. The widely accepted theoretical explanation is the propagating mass accretion rate fluctuations in the accretion flow in the context of the truncated disk model (\citealt{Ingram2016} as a review). The fluctuation is generated throughout the accretion flow and propagates inward, during which viscosity plays the key role (\citealt{Balbus1998,Ingram2011,Mushtukov2018}). The characteristic frequency of the PDS of the fluctuations is set by the local viscous frequency which is a function of the radius (Eq. (4) in \citealt{Ingram2016}). In other words, fluctuations with different characteristic frequencies come from different regions. In general, rapid variability (characterized by high frequency) is thought to be produced in the vicinity of the central object, and slow variability (characterized by low frequency) is produced in the outer parts of the accretion flow (\citealt{Ingram2016}). 

In this work, the result that the PDS of each state is composed of two different components gives us a clue that the fluctuations may be generated and propagating in two different regions (\citealt{Axelsson2018, Rapisarda2017}). We address the evidences and interpretations of the results that lead us to envisage a disk-coronal evolution picture in the following paragraphs.

In LHS, Lor$_1$ and Lor$_2$ have nearly constant broken frequencies (left panel of Figure \ref{fig:PDScompparfig}), corresponding to two different regions with different distances. Both of the variation spectra of these two regions have thermal origins (left panel of Figure \ref{fig:RMSfit}), indicating that they are close to the disk. We speculate that the corona overlaps with the disk, resulting in the overlapping regions constitute the variation.

In IMS, the wide distribution of the broken frequency of Lor (middle bottom panel of Figure \ref{fig:PDScompparfig}) reflects a wide spatial distribution of the corona. The positive correlation of broken frequency with photon energy indicates the anti-correlation of energy with distance, the smaller the distance, the higher the photon energy (Eq. (3) in \citealt{Ingram2016}). The thermal origin of the variation spectrum (middle panel of Figure \ref{fig:RMSfit}) indicates that this region is a radially distributed region close to the disk. The photon index of PL (middle upper panel of Figure \ref{fig:PDScompparfig}) reflects that both of the lower energy radiation at large distance and higher energy radiation at small distance constitute the noise, which means a radial distribution of this region. However, the variation spectrum with the absence of thermal origin (middle panel of Figure \ref{fig:RMSfit}) indicates that this region is far away form the disk. As a result, the corona of IMS has both radial and vertical distribution.

In HSS, both of the roughly constant values of broken frequency of Lor and the photon index of PL indicate that the corona is concentrated with no radial distribution. Considering the absence of thermal origin in the variation spectra, we speculate that the corona retains vertical distribution to form the jet-like geometry.

\begin{figure*}
\centering
\includegraphics[scale=0.15]{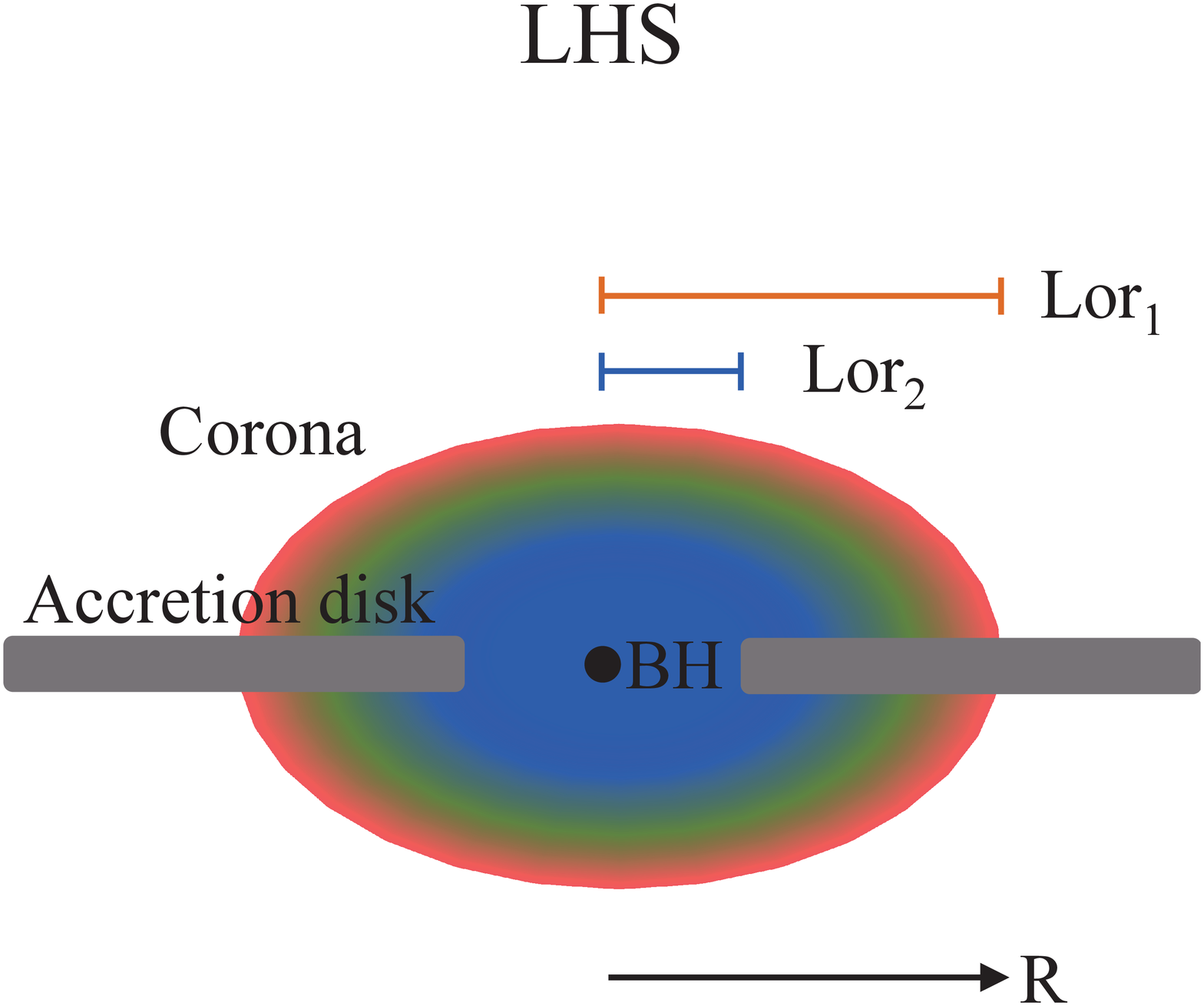}
\hspace{10 mm}
\includegraphics[scale=0.15]{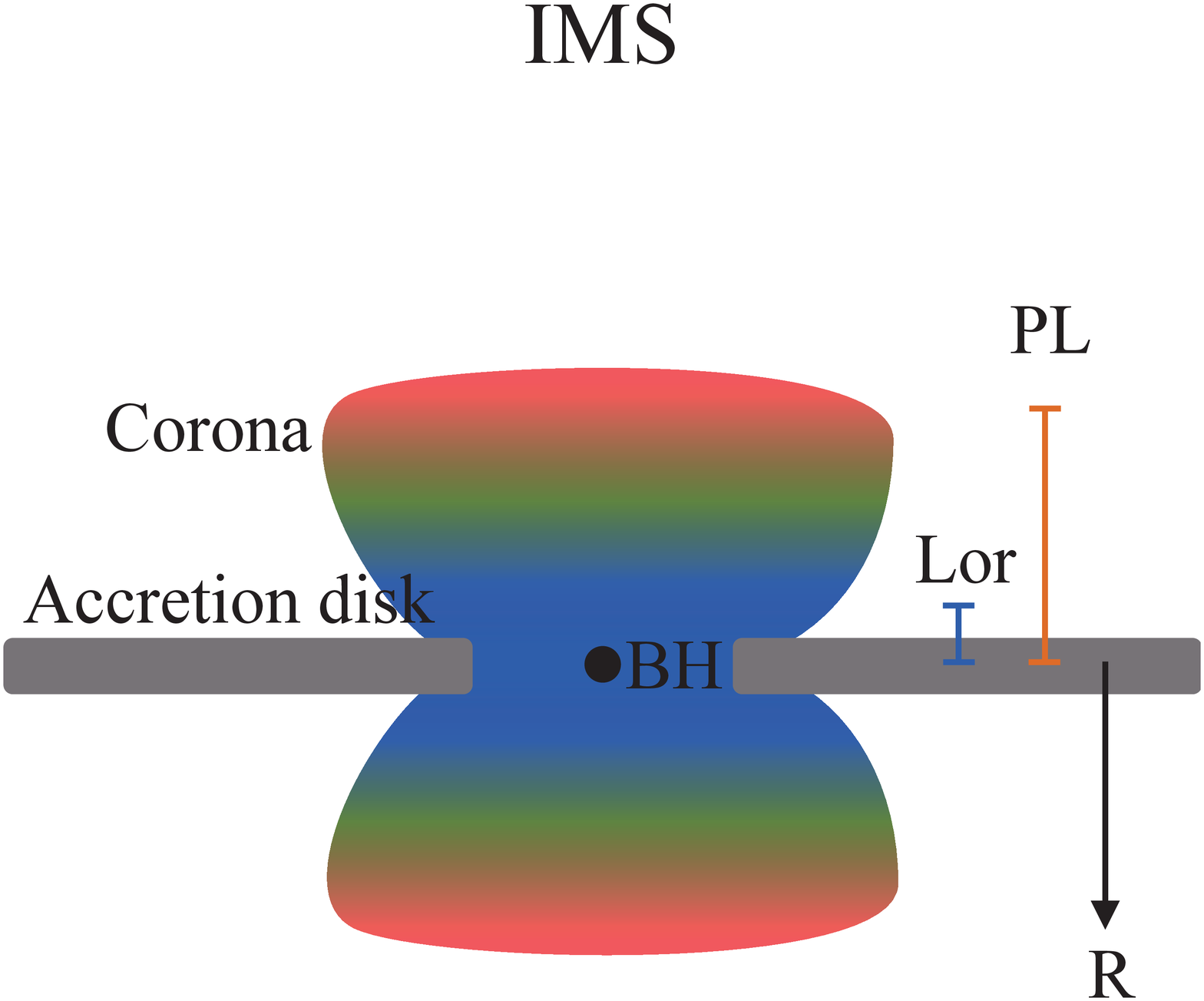}
\hspace{10 mm}
\includegraphics[scale=0.15]{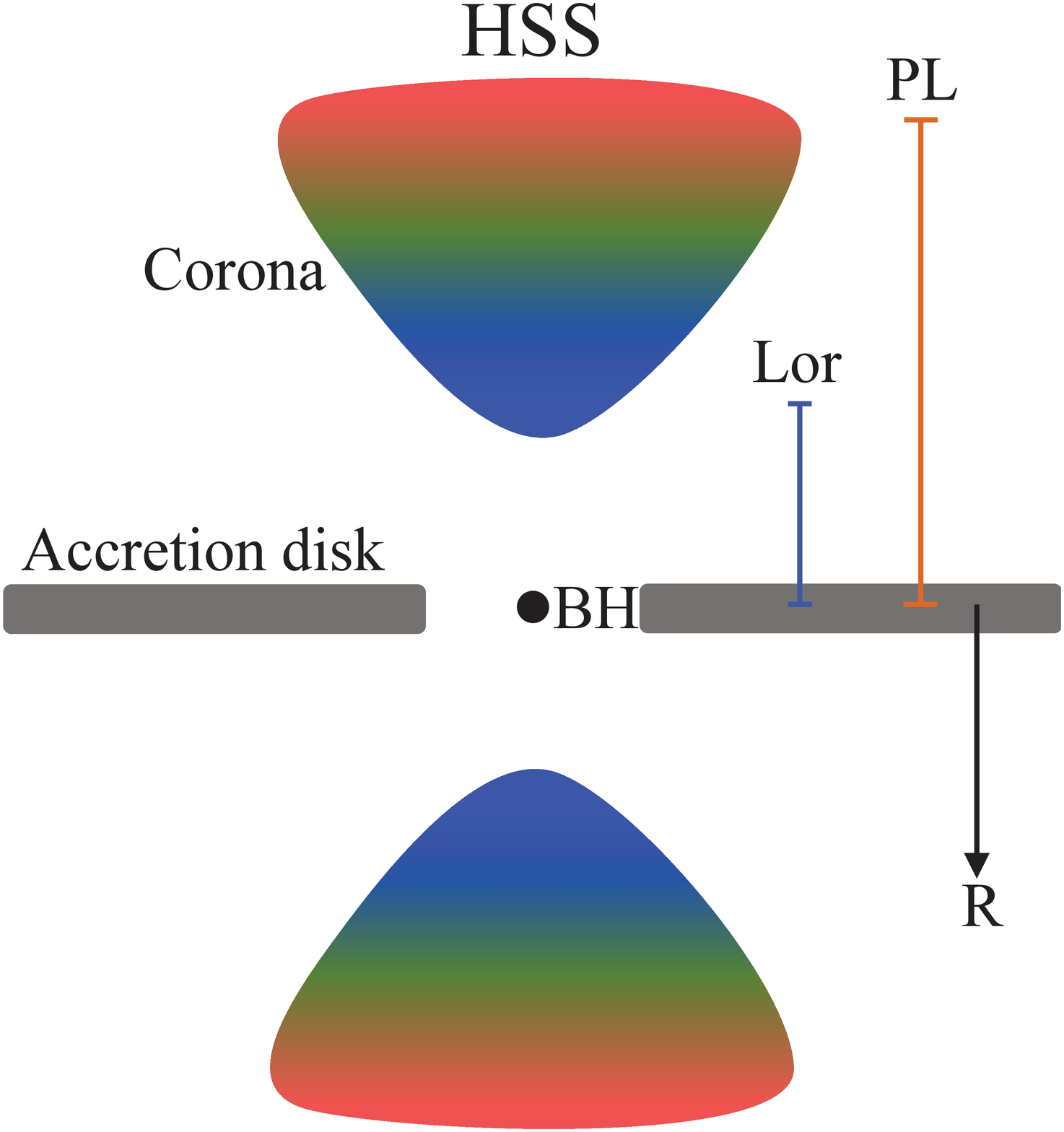}
\caption{Cartoon pictures demonstrating the disk-coronal evolution for three spectral states. The truncated inner disk edge moves nearer to the black hole as the spectrum softens. The corona wraps up the disk to form a sandwich geometry in LHS, and then gradually moves away from the disk in direction that is perpendicular to the disk until forming a jet-like geometry in HSS. The colors of corona indicate the photon energy with the higher energy in blue and lower energy in red. The relative distances of the regions that contribute to X-ray variation are indicated by lines in blue and orange.}
\label{fig:cartoonfig}
\end{figure*}

We suggest a scenario with different corona geometry for different state based on the truncated disk geometry, as illustrated in Figure \ref{fig:cartoonfig} with the relative distances of the PDSs' components marked on it. The key point is that the corona is supposed to be inhomogeneous with certain morphology and distance from the central BH for each state. It wraps up the disk to form a sandwich geometry in LHS, and then gradually moves away from the disk in direction that is perpendicular to the disk until forming a jet-like geometry in HSS. The truncated inner disk edge moves nearer to the BH as the source softens, assuming \textit R$_{\rm in}$ in Figure \ref{fig:SpecPara} indicates the true inner radius. The variation originated from the evolving corona together with the disk variation constitute the X-ray variations in Cyg X-1. 

Previous works have revealed the reduction in the spatial extent of the corona as the source softens in low mass XRBs such as MAXI J1535-571 (\citealt{Kong2020}) and MAXI J1820+070 (\citealt{Kara2019}). However, our result is opposite to this scenario because the spatial distances of the regions increase as the source softens. We speculate that the coronal evolution of Cyg X-1 is different from that of low mass XRBs. 

HSS of most low mass XRBs is characterized by a much reduced noise level than LHS because it is disk dominated, and the limited variation comes mainly from the hard spectral component (\citealt{Ingram2016}). Cyg X-1 is different not only in showing large amplitude of total variability in HSS, but with the hard spectral component having fairly large variation, which is probably due to the jet-like geometry.

As for the accretion disk, the spectral fitting results reveal that it approaches the central BH as the source softens (Figure \ref{fig:SpecPara}), while the timing analysis shows that its variability disappears as the source softens (Figure \ref{fig:RMSfit}). Our scenario suggests that the corona moves far away from the disk in vertical direction which results in the spatial separation as well as the disappearance of interaction with the disk. We might argue that the accretion disk contributes to X-ray variation along with the inward movement of the inner edge until it reaches almost to the innermost stable circular orbit (ISCO).

\begin{figure*}
\centering
\includegraphics[scale=0.18]{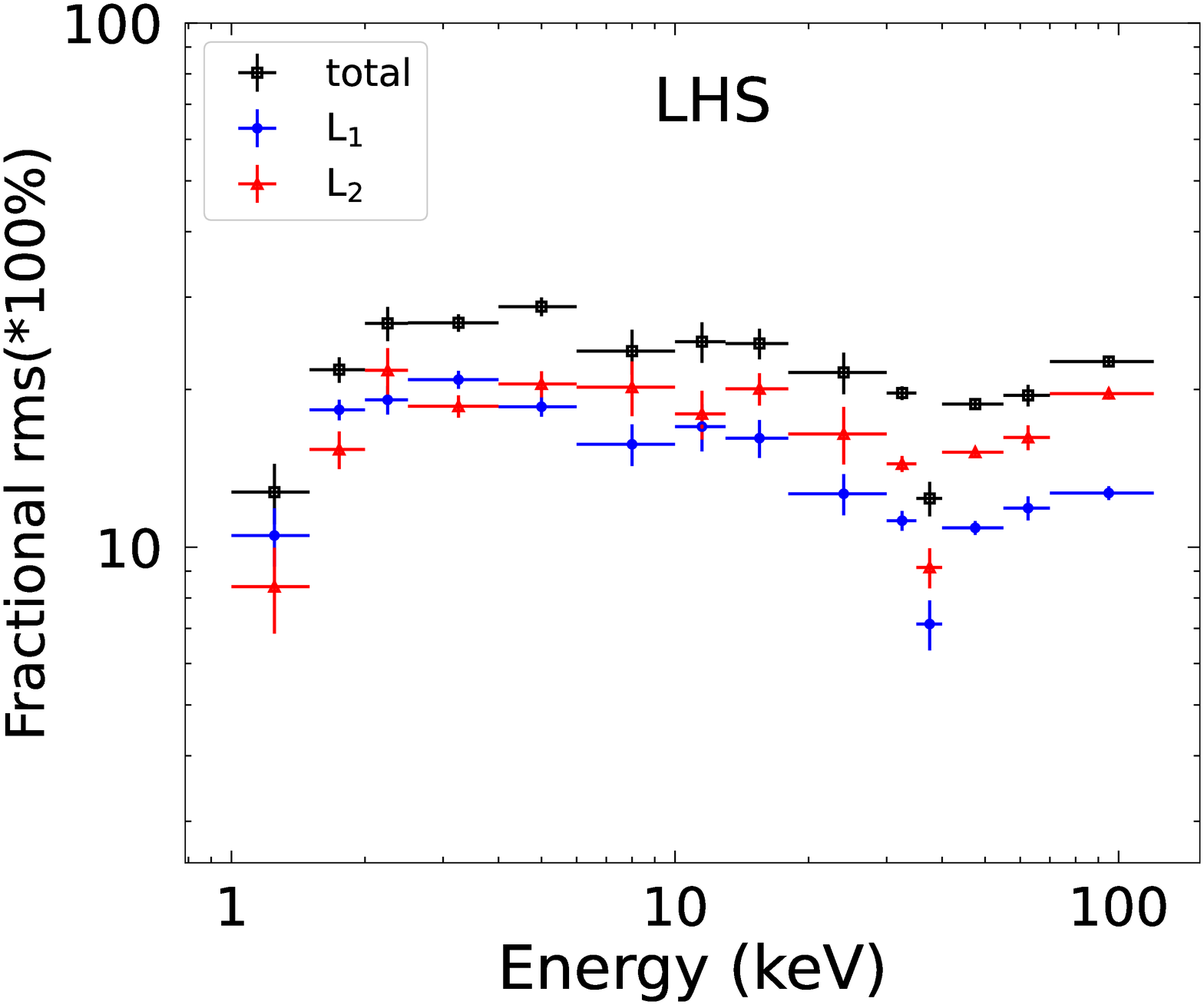}
\includegraphics[scale=0.18]{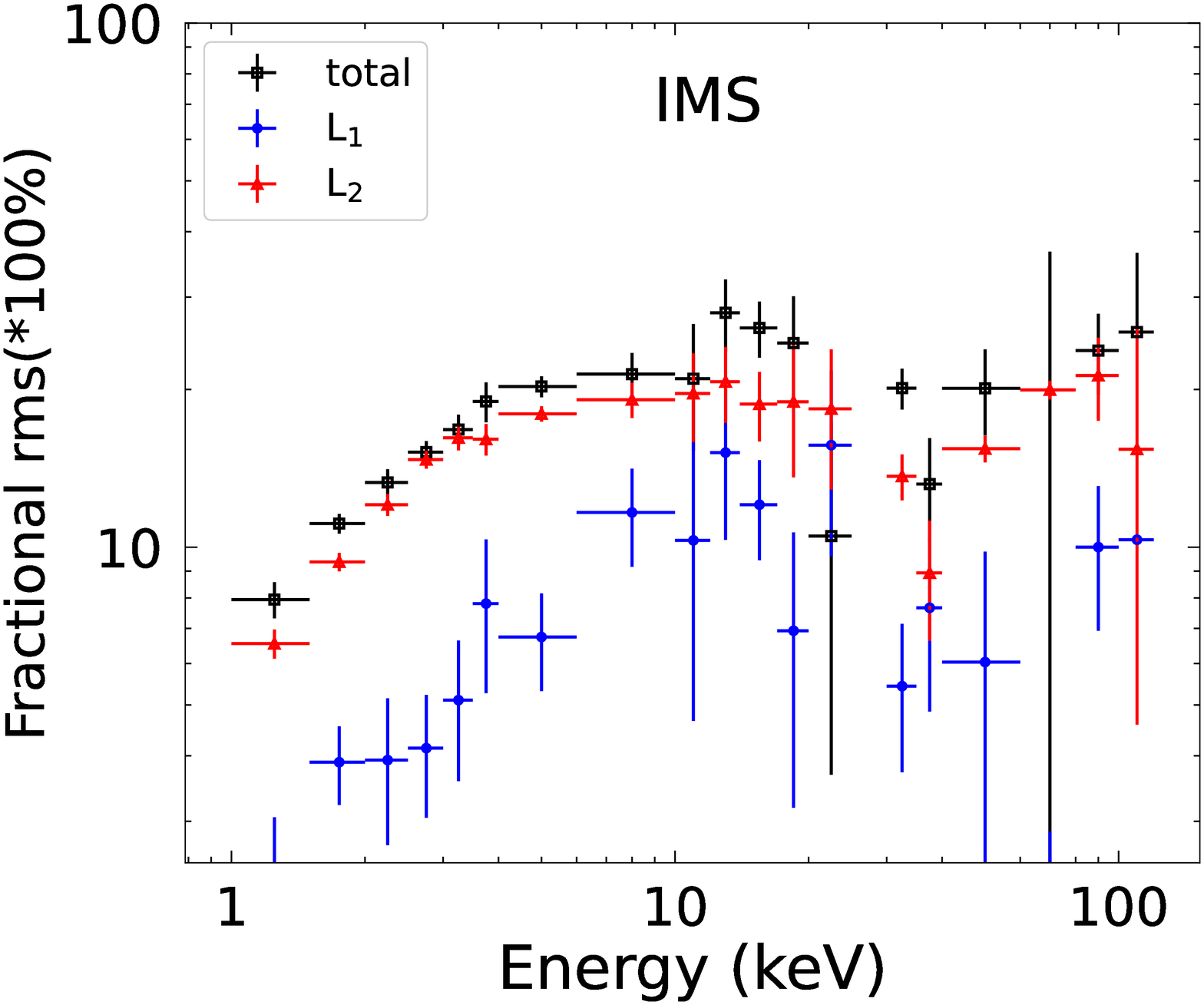}
\includegraphics[scale=0.18]{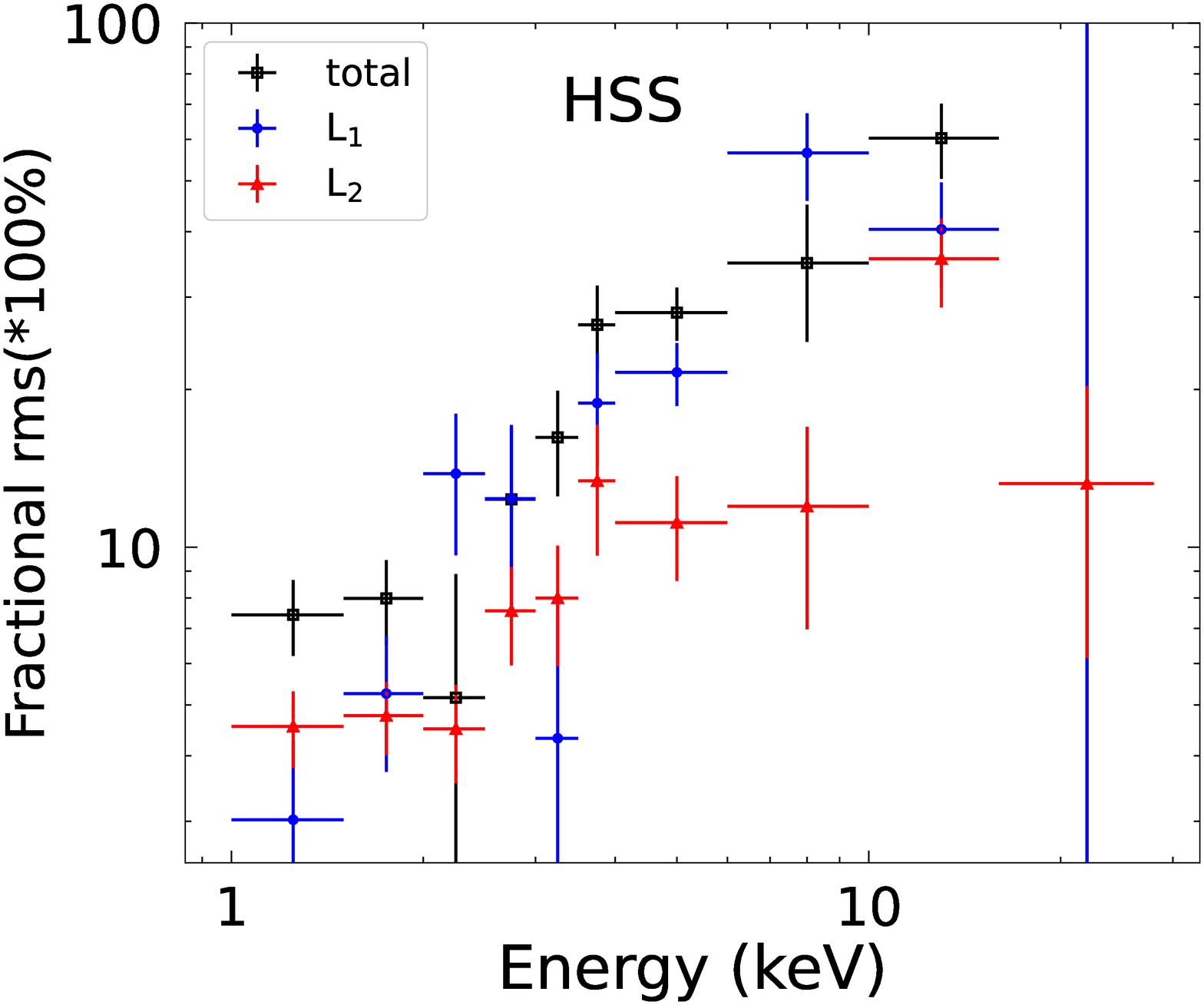}
\caption{\label{fig:simulatedRMSspec}The simulated fractional rms amplitude of the power-law component of the three states.}
\end{figure*}

The X-ray variation of corona is composed of two parts, one of which is caused by seed photon fluctuations from the cold disk and the other is its intrinsic variation. Due to the absence of disk origin in the variation spectra of HSS, the rms of the intrinsic variation of the corona probably takes the shape as shown in the right panel of Figure \ref{fig:RMSspec} (rms of HSS) and reaches its maximum amplitude in higher energy range. To check for the other two states, we use the \textit {fakeit} method in XSPEC to simulate observations of the power-law component after subtracting the \textit {diskbb} component in the RMS spectra, and then convert the simulated RMS spectra to rms as illustrated in Figure \ref{fig:simulatedRMSspec} using the reversing process of the procedure mentioned in Section \ref{RMS}. The simulated rms of HSS is also shown in the right panel of Figure \ref{fig:simulatedRMSspec} as a comparison. We may draw a conclusion that the rms of the intrinsic variation of the corona increases with energy and then stays almost stable in higher energy range, which confirms the result of the QPO rms in \citealt{Kong2020}, total rms spectra in \citealt{Gierlinski2005} and simulation results in \citealt{You2018}. Since the shape and spatial movement of the corona and the BH mass of \citealt{Kong2020} are different from those in this work, we argue that Figure \ref{fig:simulatedRMSspec} illustrates the intrinsic property of the non-thermally originated fluctuations.

\section{Conclusion} \label{conclu}

We have presented a detailed spectral-timing analysis of Cyg X-1 with broad energy coverage observations by \hxmt{}. We studied the energy spectra, PDS, the fractional rms, and the FFC resolved spectra in representative LHS, IMS, and HSS. Our main results are listed as follows:
\begin{enumerate}
    \item The inner edge of the accretion disk stays within \textit R$_{\rm g}$ and moves closer to the central BH as the source softens. 
    \item Spectral states differ in different PDS components: Two Lorentzians with distinct broken frequencies in LHS; A power-law component appears with only one of the Lorentzians left in IMS; The power-law becomes stronger, and the broken frequency of the Lorentzian shifts to lower band in HSS.
    \item The rms slightly decreases with increasing photon energy in LHS, and rises with energy monotonically to the maximum value and then stays roughly constant in IMS and HSS.
    \item The variable part of the X-ray flux have both thermal and non-thermal origins in harder states. As the source softens, the thermal origin disappears.
\end{enumerate}

Our results prompt a speculation of the evolving corona along with state transitions rather than an evolving, turbulent accretion disk. We argue that the major role that contributes to the X-ray variation is the hot corona rather than the accretion disk. We suggest a scenario with different corona geometry for each state based on the truncated disk geometry in which the corona wraps up the disk to form a sandwich geometry in LHS, and then gradually moves away from the disk in direction that is perpendicular to the disk until forming a jet-like geometry in HSS. The coronal evolution that is distinguishing from other XRBs indicates the particularity of Cyg X-1 as a HMXRB.
 
\begin{acknowledgments}

This work made use of data from the \textit{Insight}{\rm -HXMT} mission, a project funded by the China National Space Administration (CNSA) and the Chinese Academy of Sciences (CAS),as well as the MAXI data provided by RIKEN, JAXA and the MAXI team. We gratefully acknowledge the financial support from the National Key R\&D Program of China (2021YFA0718500), the National Natural Science Foundation of China (Grant No. U1838201, U1838202, 11733009, U2031205, 11961141013), the Joint Research Fund in Astronomy under the cooperative agreement between the National Natural Science Foundation of China and the Chinese Academy of Sciences (Grant No. U1631242) and the National Basic Research Program (973 Program) of China (Grant No. 2014CB845800). JL thanks the support from the National Natural Science Foundation of China grants No. 12173103, U2038101, U1938103, 11733009, and the Guangdong Major Project of the Basic and Applied Basic Research grant 2019B030302001.

\end{acknowledgments}

\bibliographystyle{aasjournal}
\bibliography{ref}{}


\end{document}